\documentclass[12pt]{article}
\pdfoutput=1
\usepackage{jheppub}
\usepackage{soul}
\usepackage{setspace}
\usepackage{slashed} 
\usepackage{graphicx,url}
\usepackage{booktabs}
\usepackage{bbm}
\usepackage{braket}
\usepackage{colortbl}
\usepackage{amsmath,hyperref,amssymb,cancel,stmaryrd}
\usepackage{amsfonts}
\usepackage{slashed}
\usepackage{mathrsfs}
\usepackage{verbatim} 
\usepackage{caption}
\usepackage[usenames,dvipsnames]{xcolor}
\usepackage{color} 
  
\usepackage{epsfig}

\def\Or[#1]{{\text{O}}\left({#1}\right)}
\def\dotl[#1,#2]{\left\langle #1, #2 \right\rangle}
\def\dotlb[#1,#2]{[ #1, #2 ]}
\def\dotp[#1,#2]{(#1) \cdot (#2)}
\def\aff[#1,#2]{\hat{#1}(#2)}
\def\n4sym{{\cal N}=4 SYM}
\def\>{\rangle}
\def\<{\langle}

\def\weight[#1,#2,#3]{\{(#1),#2,#3\}}
\def\ads[#1]{$\text{AdS}_{#1}$}

\newcommand{\ba}{\begin{eqnarray}}
\newcommand{\ea}{\end{eqnarray}}
\newcommand{\bal}{\begin{aligned}}
\newcommand{\eal}{\end{aligned}}
\newcommand{\be}{\begin{eqnarray}}
\newcommand{\ee}{\end{eqnarray}}
\newcommand{\bq}{\begin{equation}}
\newcommand{\eq}{\end{equation}}
\newcommand{\benn}{\begin{equation*}}
\newcommand{\eenn}{\end{equation*}}
\newcommand{\bi}{\begin{itemize}}  
\newcommand{\ei}{\end{itemize}}
\renewcommand{\d}{\partial}

\newcommand{\Mcal}{{\mathcal M}}

\newcommand{\Ocal}{{\mathcal O}}
\newcommand{\cO}{{\mathcal O}}

\newcommand{\CL}{{\cal L}}

\newcommand{\CM}{{\cal M}}

\newcommand{\CO}{{\cal O}}

\newcommand{\nn}{\nonumber}
\renewcommand{\Im}[0]{\operatorname{Im}}

\newcommand\oo\infty
\newcommand\s\sigma
\newcommand\de\delta
\newcommand\De\Delta
\newcommand{\p}{\partial}
\newcommand\f\phi
\newcommand\g\gamma
\newcommand\x\times

\newcommand{\fr}{\frac}

\newcommand{\Dmax}{\De_{\max}}
\newcommand{\dmax}{\De_{\max}}
\newcommand\lrpar{\raise .8ex\hbox{$^\leftrightarrow$} \hspace{-9pt}
\partial}

\newcommand{\dagg}{\dagger}

\newcommand{\Tr}{\textrm{Tr}}

\usepackage{listings}
\usepackage{xcolor}
\usepackage{mdframed}

\usepackage{afterpage}
\usepackage{bookmark}

\newenvironment{monospace}{\ttfamily}{\par}

\newcommand{\kvec}{\boldsymbol{k}}

\newcommand{\pr}[1]{\left(#1 \right)}

\newcommand{\ptl}{\partial}
\newcommand{\Kvec}{\boldsymbol{k}}

\newcommand{\Nc}{N_c}

\makeatletter
\def\@fpheader{\vspace{-.1cm}}
\makeatother

\title{
 Chiral Limit of 2d QCD Revisited with Lightcone Conformal Truncation
}

\author{Nikhil Anand$^a$, A. Liam Fitzpatrick$^b$, Emanuel Katz$^b$, and Yuan Xin$^c$}

\affiliation[a]{Department of Physics, McGill University, Montr\'eal, QC H3A 2T8, Canada}
\affiliation[b]{Department of Physics, Boston University, Boston, MA 02215, USA}
\affiliation[c]{Department of Physics, Yale University, New Haven, CT 06520, USA}

\abstract{We study the chiral limit of 2d QCD with a single quark flavor at finite $N_c$ using LCT.  By modifying the LCT basis according to the quark mass
in a manner motivated by 't Hooft's analysis, we are able to restore convergence for quark masses much smaller than the QCD strong coupling scale.  
For such small quark masses, the IR of the theory is expected to be well described by the Sine-Gordon model.  We verify that LCT numerics are able to 
capture in detail the spectrum and correlation functions of the Sine-Gordon model.  This opens up the possibility for studying deformations of various integrable CFTs 
using LCT by considering the chiral limit of QCD like theories.
}

\arxivnumber{}

\begin{document}

\maketitle  

%\tableofcontents

%%%%%%%%%%%%%%%%%%%%%%%%%%%%%%%%%%%%%%%%%%%%%%%%%%%%%%%%%%%%%%%%%%%%%%%%%%%%%
%%%%%%%%%%%%%%%%%%%%%%%%%%%%%%%%%%%%%%%%%%%%%%%%%%%%%%%%%%%%%%%%%%%%%%%%%%%%%
%%%%%%%%%%%%%%%%%%%%%%%%%%%%%%%%%%%%%%%%%%%%%%%%%%%%%%%%%%%%%%%%%%%%%%%%%%%%%

%\newpage
\section{Introduction}
\label{sec:intro}

Understanding gauge theories in the non-perturbative regime is a worthy goal for many applications to both condensed matter and in particle physics.  For the most part lattice gauge theory (LGT) has been the preferred tool to study gauge theories.  Nevertheless, LGT is not ideal when it comes to capturing real time dynamics, and faces complications in maintaining certain symmetries, such as chirality or supersymmetry.  Hamiltonian truncation methods, on the other hand, allow for direct access to time evolution, and as they do not discretize space, can often preserve some subset of the symmetries.  In the 2d context, there has been much work studying gauge theories using lightcone (LC) quantization, where one can choose lightcone gauge and still maintain gauge invariance in the presence of a hard LC momentum cutoff.  In this context, most of the work has been done using Discrete Lightcone Quantization (DLCQ)\cite{Pauli:1985ps}.

A model which remained out of reach, however, was a 2d version of real 4d QCD, where the strongly bound quarks have nonvanishing masses much lighter than the strong coupling scale (set in 2d by the gauge coupling $g$).  As we will review below, though it is possible to accommodate either strictly massless quarks or heavy quarks \cite{Pauli:1985pv, Pauli:1985ps, Hornbostel:1988fb,Hornbostel:1988ne,Bhanot:1993xp,
Demeterfi:1993rs,Anand:2020gnn,Katz:2014uoa}, for light quarks, bound state wavefunctions develop features which can be difficult to approximate.  Consequently, DLCQ  as well a generic Lightcone Conformal Truncation (LCT) approach will lead to slow convergence.  Studying QCD in the light mass, or chiral limit, is desirable not only because of its closer resemblance to its 4d cousin, but also because the effective theory describing the low energy degrees of freedom of QCD like theories can be very rich in its own right.  Indeed, recently, Delmastro, Gomis and Yu \cite{Delmastro:2021otj} have offered a full classification of the IR of QCD theories in the chiral limit, building on the earlier insights of \cite{Kutasov:1993gq, Kutasov:1994xq,Bhanot:1993xp}; examples  include minimal models and WZW models.  Hence, a method which can efficiently capture the chiral limit of QCD like theories, can also be used to study relevant deformations of a wide class of integrable CFTs.

In this paper, we take the first step towards exploring the chiral limit by considering the simplest case of 2d QCD with a single quark using LCT.  Anomaly matching requirements constrain the IR of this theory to be that of a single real scalar ``pion".  By standard arguments, a mass deformation results in a cosine potential for the pion.  This is of course the well-known Sine-Gordon model, for which many results are known from integrability.  This setting is thus ideal to check LCT and its ability to capture the $c=1$ degree of freedom.
 
We begin in section \ref{sec_SG} with a review of the connection between the chiral Lagrangian of QCD and the Sine-Gordon model in a somewhat more modern language.
In section \ref{masslessLCT}, after reviewing the lightcone formulation of QCD, we first focus on the strict massless limit, explaining why the naive LCT basis contains an exactly massless sector with $c=1$ at finite $N_c$.  Next, in section \ref{chiralLargeN} we review the challenge posed by light quarks, using the large $N_c$ limit as warmup.  In this limit, we describe a change to the LCT basis in terms of the quark mass (as motivated by 't Hooft's analysis \cite{t1993two}) which restores fast convergence.  In section \ref{finiteNbasis} we generalize the modified LCT basis to multi-particle states (essential to solve the theory at finite $N_c$).  We find that the necessary modification of the LCT basis agrees well with the earlier two-particle approximation suggested by Sugihara et al. \cite{Sugihara}, especially for very small quark masses.  Finally, in section \ref{results} we report on our numerical results for the spectrum and for correlation functions for $N_c=3$, and compare them to detailed expectations from the Sine-Gordon model.  We find very good agreement between numerics and integrability in the limit of very small quark masses.  In particular, we reproduce numerically the c-function (or stress-tensor correlation function) of the Sine-Gordon model in Fig. \ref{fig:pltCFuncSG}.
 
 \section{2d Chiral Lagrangian and Sine-Gordon}
 \label{sec_SG}
 
Before we consider 2d QCD in Hamiltonian truncation, we will first determine what the spectrum should look like in the limit of small quark mass $m_q$ by studying the chiral Lagrangian for the pion.  It turns out that in this limit, the low-energy theory is described by the Sine-Gordon model \cite{steinhardt1980}.  Here, we will rederive this result in modern terms.  We begin with the statement that the axial symmetry of the theory is nonlinearly realized by shifts on the pion.  That is,
 \be
 \bar{q} q \sim e^{i \pi/f_\pi},
 \ee
 so the axial symmetry $q \rightarrow e^{i \alpha \gamma_5} q$ acts on $\pi$ by $\pi \rightarrow \pi+2\alpha f_\pi$.  Therefore, at $m_q=0$, the chiral Lagrangian is simply $\CL = \frac{1}{2} (\partial \pi)^2$, plus irrelevant interactions.  When $m_q$ is nonzero but small, we can treat it as a spurion for the axial symmetry and add the following invariant combination to the action:
 \be
 \CL \supset b \left( m_q \Lambda e^{ i \pi /f_\pi}+ m_q^* \Lambda e^{- i \pi /f_\pi} \right) , 
 \ee
 where $\Lambda$ is the confinement scale, and $b$ is an unknown dimensionless constant.
 
 Remarkably, in 2d we can fix $f_\pi$ in terms of $N_c$ using anomaly matching.  In terms of the pion field, the axial current is
 \be
 J_A^\mu =2 f_\pi \partial_\mu \pi,
 \ee
 so that $[ i \int d x J_A^0(x) , \pi(y)] = 2f_\pi$. To relate $f_\pi$ and $N_c$, we compute the axial anomaly coefficient in the IR and in the UV.  In the IR at zero $m_q$,
 \be
 \< J_A^\mu(p) J_A^\nu(-p)\> =4 f_\pi^2 \frac{p_\mu p_\nu}{p^2},
 \ee
 whereas in the UV the quark loop contributes
 \be
 \<J_A^\mu(p) J_A^\nu(-p)\> \supset \frac{N_c}{\pi} \frac{p_\mu p_\nu}{p^2}. 
 \ee
 Matching the UV and IR anomaly coefficient, we obtain the relation
 \be
 f_\pi^2 = \frac{N_c}{4 \pi}.
 \ee
Putting everything together and taking $m_q$ to be real, we identify the chiral Lagrangian as the sine-Gordon model,
\be\label{eq:SG-Lagrangian}
\CL = \frac{1}{2} (\partial \pi)^2 + \lambda \cos \beta \pi,
\ee
with
\be
 \lambda = b m_q \Lambda, \qquad \beta = 2\sqrt{\frac{\pi}{N_c}}.
 \ee
 It is well-known that due to quantum effects, the scaling dimension of the cosine operator is
 \be
 \Delta[ \cos \beta \pi ] = \frac{\beta^2}{4\pi} = \frac{1}{N_c},
 \label{eq:CosineDimension}
 \ee
 and so is relevant for physical values of $N_c$.  There are also subleading interactions with additional powers of $m_q$, the next one being $\sim m_q^2 \cos(2 \pi/f_\pi)$.  By equation (\ref{eq:CosineDimension}), the dimension of this subleading interaction is $\Delta = \frac{4}{N_c}$.
 
 The sine-Gordon model in 2d is integrable, and the mass spectrum of states is known in closed form \cite{Dashen:1975hd}.  Our main interest will be in a handful of the lightest states.  These are a baryon, a meson, and a bound state of two mesons, with masses $m_B, m_M$ and $m_2$ respectively.  They are related to each other by
\be
m_M = 2 m_B \sin \left( \frac{\pi}{2(2N_c-1)}\right), \qquad m_2 = 2 m_B \sin \left(  \frac{\pi}{2N_c-1}\right) . 
\label{eq:SineGordonMassPrediction}
\ee 
Expanding in $1/N_c$, 
\be
\frac{m_2}{2m_M} \approx 1  - \frac{ \pi^2 }{32 N_c^2}  + \CO(N_c^{-3}),
\ee
so the bound state mass is very close to $2m_M$ even for modestly large $N_c$; in particular, at $N_c=3 \ [N_c=2]$, $\frac{m_2}{2 m_M} = 0.95 \ [0.87]$.

 \section{2d QCD in Lightcone Hamiltonian Truncation}
\label{masslessLCT}

\subsection{Lightcone Hamiltonian}

 Our main focus in this work is to apply LCT to two-dimensional QCD with fundamental matter, given by the Lagrangian \begin{equation}
	\mathcal{L} = -\frac{1}{2} \textrm{Tr} F^{\mu\nu} F_{\mu\nu} + \overline{\Psi}(i \slashed{D} - m_q) \Psi.
\end{equation} Our goal is to find the eigenvalues and eigenvectors of the invariant mass-squared operator $M^2 = 2P_+ P_-$, working in lightcone quantization,  i.e. quantizing on surfaces of constant lightcone ``time'' $x^+$ where \begin{equation}
	x^{\pm} \equiv \frac{x^0 \pm x^1}{\sqrt{2}}.
\end{equation} 
A particularly advantageous gauge choice is lightcone gauge, $A_-=0$.  The Hamiltonian $P_+$  in this gauge is straightforward to derive by integrating out the nondynamical degrees of freedom. Briefly, one separates the fermion field $\Psi$ into left- and right-movers
 \begin{equation}
	\Psi_i = \frac{1}{2^{1/4}}\begin{pmatrix}
	\psi_i \\ \chi_i
\end{pmatrix}
\end{equation} where $i$ is the fundamental index which we will suppress from now on to avoid clutter. The Lagrangian then takes the form \begin{equation}
	\mathcal{L} = i(\psi^\dagg \ptl_+ \psi + \chi^\dagg \ptl_- \chi) + g \psi^\dagg A_+^A T^A \psi + \Tr(\ptl_- A_+)^2 -\frac{m_q}{\sqrt{2}}(\chi^\dagg \psi + \psi^\dagg \chi).
\end{equation} 
The right mover $\chi$ and the gauge field $A_+$ are non-dynamical, and can be integrated out to obtain the following form of the Hamiltonian: \begin{equation}
	P_+ = \int dx^- T_{-+} =  \int dx \left[\frac{m_q^2}{2} \mathrm{:} \psi^\dagg \frac{1}{i\ptl}\psi \mathrm{:} - \frac{g^2}{2} \mathrm{:}\psi^\dagg T^A \psi \mathrm{:} \frac{1}{\ptl^2} \mathrm{:} \psi^\dagg T^A \psi \mathrm{:} \right], \label{eqn:2dqcdhamiltonian}
\end{equation} where we have suppressed $-$ subscripts on the derivatives (which will be  implicit from now on) and the superscript on the spatial coordinate $x^-$.

\subsection{Massless Quarks Review}

The presence of a small but nonzero quark mass $m_q \ll g$ presents challenges that will be the main focus of this work.  To warm up, we first review the massless case, and describe the advantages of the LCT approach here. 

At $m_q=0$, the only interaction is the quartic term in (\ref{eqn:2dqcdhamiltonian}).  The gauge boson propagator $\frac{1}{\partial^2}$ is ambiguous and must be defined more precisely; 't Hooft \cite{t1993two} defined it in momentum space as the following `principal value prescription':
\be
\frac{1}{\partial^2} \rightarrow P.V. \frac{1}{k_-^2} \equiv \frac{1}{2}\left( \frac{1}{(k_-+ i \epsilon)^2}+ \frac{1}{(k_-- i \epsilon)^2}\right).
\label{eq:PV}
\ee
This definition is manifestly finite (eg  by contour deformation) when integrated against any smooth function. 
 In appendix \ref{app:GaugeHeff}, we show that 't Hooft's principal value condition for the nonlocal interaction can be derived by using the lightcone effective Hamiltonian $H_{\rm eff}$ from \cite{Fitzpatrick:2018ttk}.

 The interaction as written in (\ref{eqn:2dqcdhamiltonian}) is not normal-ordered in the usual sense, but for convenience it can be separated  into a normal-ordered operator and  the contribution from the internal contractions of the fermions.  These internal contractions are easily evaluated and produce a finite shift in the fermion mass term operator, $\sim \psi^\dagger \frac{1}{\partial} \psi$.
\be
\mathbf{:}\psi^\dagger T^A \psi \mathbf{:} \frac{1}{\partial^2} \mathbf{:} \psi^\dagger T^A \psi \mathbf{:} \cong \mathbf{:} \psi^\dagger T^A \psi \frac{1}{\partial^2} \psi^\dagger T^A \psi\mathbf{:} - \frac{C_2(N_c)}{\pi}\mathbf{:}\psi^\dagger \frac{1}{\partial} \psi\mathbf{:} 
\label{eq:InteractionNormed}
\ee
where $C_2(N_c)= \frac{N_c^2-1}{2N_c}$. An important effect of the second term in (\ref{eq:InteractionNormed}) is that the the state created by the axial current remains exactly massless when the bare quark mass $m_q$ vanishes, due to a cancellation between the two terms in (\ref{eq:InteractionNormed}).  One can also reverse this logic, and use the fact that anomaly-matching requires the current to create a massless state in order to fix the coefficient of the second term. 

Although representing the interaction as the sum of normal-ordered terms in (\ref{eq:InteractionNormed}) is convenient for the Fock space representation of the states, it obscures the conformal structure of the interaction and in particular the fact that the dynamics of the interacting theory are controlled by the current algebra of the theory and its representations.  This fact and some of its implications were explained in \cite{Kutasov:1994xq} and have been exploited many times since (eg \cite{Dempsey:2021xpf}).  It is especially advantageous in LCT, where the main point of the LCT basis is that all basis states are simply Fourier transforms of local CFT primary operators acting on the vacuum:
\begin{equation}
	\ket{\cO_i, p} \equiv \frac{1}{N_{\cO_i}} \int d^d x e^{-i p \cdot x} \cO_i(x) \ket{\textrm{vac}}, \label{eq:lctbasis}
\end{equation}
 with $N_{\cO_i}$  a normalization factor.  This connection between the states and primary operators allows one to take advantage of the conformal structure of the UV fixed point.  More explicitly, the basis states take the form
\cite{Katz:2014uoa} \begin{equation}
\label{eq:qcdMonomial}
	\begin{aligned}
		\ket{\Ocal,p} &= \frac{1}{N_{\Ocal}}\int  dx \, e^{-i p x} \cO(x)\\
		&=  \frac{1}{N_{\Ocal}}\int  dx \, e^{-i p x} \sum_{\Kvec} C_{\Kvec} \left( \ptl^{k_{11}} \psi_{i_1}^\dagg \ptl^{k_{21}} \psi_{i_1} \right) \left( \ptl^{k_{12}} \psi_{i_2}^\dagg \ptl^{k_{22}} \psi_{i_2} \right) \dotsb \left( \ptl^{k_{1n}} \psi_{i_n}^\dagg \ptl^{k_{2n}} \psi_{i_n} \right)(x) \ket{\textrm{vac}}, 
	\end{aligned}
\end{equation} where $N_{\cO}$ is a normalization factor and where in the second line, we have expanded the operator $\cO(x)$ in terms of its ``monomial'' constituents.  
The $C_{\Kvec}$ coefficients are chosen such that the states~\eqref{eq:qcdMonomial} form a complete, orthogonal basis.\footnote{In Fock space language, the conformal operator basis used in LCT simply corresponds to taking multi-quark states with wavefunctions that are polynomials in momentum for the individual quark components.  For example,  a general two-quark singlet state can be written as
\be
|\Psi, p_-\>_2 \equiv  \int_0^{p_-}  dq_- f(q_-) | q_- , a; p_- - q_-, a\>.
\ee
summation over $a=1, \dots, N$ implied.  The massless pion is just a two-quark singlet with a constant wavefunction, $f(q_-) = \textrm{const}$. }

To see how the spectrum is controlled by the Kac-Moody (KM) current algebra, recall first that in the UV, the $N_c$ fermions have a $U(N_c)_1$ current algebra, and its $SU(N_c)_1$ subgroup is gauged.  We can write the $U(1)$ current in $U(N_c)_1$ as $J^0 \sim \psi_i^\dagger \psi_i$, and the $SU(N_c)_1$ currents as $J^a \sim \psi_i^\dagger T^a_{ij} \psi_j$.  All operators with vanishing baryon number can be constructed out of products of $J^a, J^0$ and their derivatives.\footnote{For instance, the stress tensor is $T= \frac{1}{2} (\partial \psi^\dagger_i \psi_i - \psi_i^\dagger \partial \psi_i) = \frac{1}{2N_c} (J^0 J^0)(z) + \frac{1}{2(N_c+1)} (J^a J^a)(z)$.} In this context, it is more convenient to define composite operators by taking contour integrals,  ie by using the product $(AB)(z) \equiv \oint \frac{dw}{2\pi i } \frac{A(z)B(w)}{z-w}$. The reason this is more convenient is that now the matrix elements of the Hamiltonian are all manifestly just integrals of correlators of currents, and correlators of currents are completely fixed by the algebra. 

 The massless sector is particularly simple in this description: the massless states are just those made from products of $J^0$ and its derivatives.   To see why, note that the interaction $J^a \frac{1}{\partial^2} J^a$ is simply a (double) integral over two insertions of the $SU(N_c)_1$ currents, so we can extract the Hamiltonian matrix elements from correlators with an insertion of $J^a(y) J^a(y')$.  By holomorphicity, correlators of currents are fully determined by their poles, so a correlator of the form $\< \CO(x) J^a(y) J^a(y') \CO'(z)\>$ vanishes if it has  no poles as a function of $y'$.\footnote{The contraction of $J^a$ with itself, $J^a(y) J^a(y') \sim \frac{N_c^2-1}{(y-y')^2}$, is a bubble diagram that is removed in lightcone quanitzation.} Therefore, states made from primary operators $\CO$ without singular terms in the $J^a(y) \CO(0)$ OPE are massless.   In particular,  correlators of the form $\< J^0(z_1) \dots J^0(z_n) J^a(y) J^a(y')\>$ vanish.  Since the matrix elements of states built from powers of $J^0$ are just integrals over such correlators, they also vanish, and the massless states in LCT are simply the primaries that can be constructed out of $J^0(z)$, without using $J^a$.

Returning to the massive sector of the theory, we can ask how to identify the massive single particle states.  This is not as simple as it might sound since massive bound states appear in the midst of  a multi-particle continuum due to the pions, and therefore do not particularly stand out if one looks at the spectrum alone.   A useful way to identify the bound states is to look at the spectral density of the stress tensor.  Due to anomaly matching, massless particles in 2d cannot have interactions and only contribute to $T_{--}$ through a kinetic term in the IR that looks like $\sim (\ptl \pi)^2$. Thus, $T_{--}$ only has overlap with  massive bound states, and the lightest massive single-particle mesons appear as isolated poles in the spectral density. At higher energies, $T_{--}$ also has overlap with the multi-particle continuum of these massive particles. 

\begin{figure}[htbp]
\centering
\includegraphics[width=.8\linewidth]{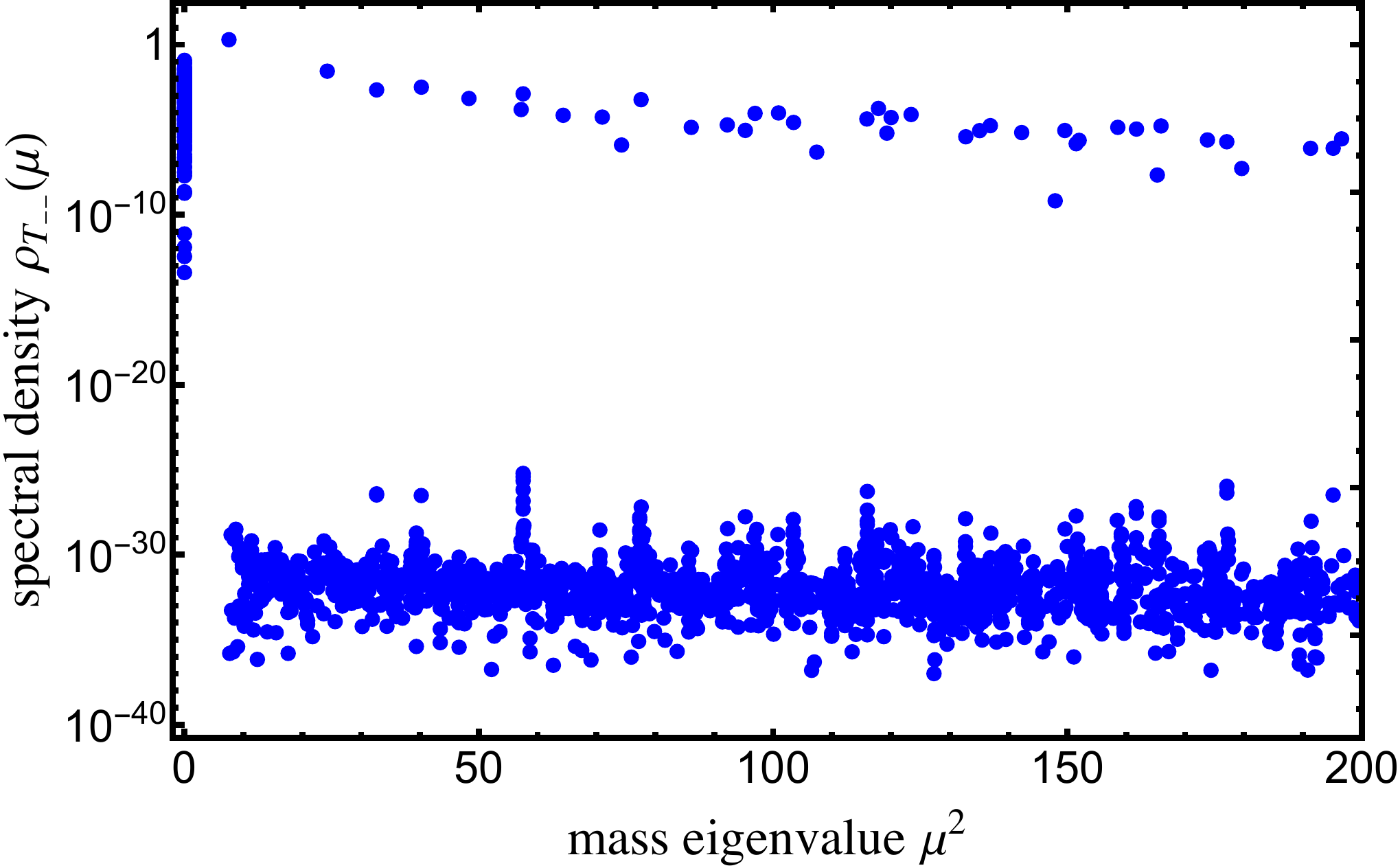}
\caption{\label{fig:TmmSpecDensity}Shows the spectral density of $T_{--}$ at a truncation cutoff of $\Dmax = 15$, which corresponds to 3,032 basis states at $N_c = 3$. The massless sector does not contribute to the spectral density due to anomaly matching. Massive states do contribute to the spectral density and thus appear as isolated poles above this sea of states with massless particles. We use units where $g^2 N_c/\pi =1$.
}
\end{figure}

 Concretely, the spectral density of $T_{--}$ can be obtained by summing over overlaps 
 \begin{equation}\label{eq:spectralDensityDefinition}
	\rho_{T_{--}}(\mu) = \sum_j | \langle T_{--}(0) | \mu_j, p \rangle |^2 \delta(\mu^2 - \mu_j^2).
\end{equation} 
For example, Fig.~\ref{fig:TmmSpecDensity} shows the spectral density of the $T_{--}$ component of the stress-energy tensor at $\Dmax = 15$ which corresponds to  3,032 basis states for $N_c = 3$. 
This is shown in Fig.~\ref{fig:TmmSpecDensity}, where we can read off the first few massive meson states.

\section{The Challenge at Small Quark Mass}
\label{chiralLargeN}
\subsection{Small Quark Mass and Boundary Conditions}

In this section we will discuss the regime where the fermion mass term
\begin{equation}
\label{eq:ferimon-mass}
\delta\CL_{\rm mass} = -m_q^2 \psi^\dagger \frac{1}{\p} \psi
= \int \frac{dp}{8\pi} (a_p^\dagger a_p + b_p^\dagger b_p) \frac{m_q^2}{2p}
\end{equation}
is small but nonzero. 

The presence of a quark mass term forces one to adjust the basis of states to accommodate a new boundary condition for the wavefunctions.  To see this, consider the mass term matrix element of
an arbitrary 2-particle state
\begin{equation}
\begin{aligned}
\label{eq:IRDiv}
\< \d^{k_1}\psi^\dagger \d^{k_2}\psi, p | M | \d^{k_1'}\psi^\dagger \d^{k_2'}\psi, p' \> &= 
(2\pi)\delta(p-p') \times \\
&
\int \frac{dp_1}{8\pi} 
	p_1^{k_1+k_1'} p_2^{k_2+k_2'} 
 \pr{\frac{m_q^2}{2p_1}+\frac{m_q^2}{2p_2}}_{p_2=p-p_1} \, .
\end{aligned}
\end{equation}
Due to the $\frac{1}{p_i}$ factors, the integral is divergent at the $p_i\rightarrow 0$ or $p_i \rightarrow p$
boundaries, if either $k_1=k_1'=0$ or $k_2=k_2'=0$. The consequence of this IR
divergence is that some states become infinitely massive and are lifted 
out of the IR spectrum. The states that remain in the spectrum are those with wavefunctions that vanish at $p_i \rightarrow 0$ for any of the individual fermion momenta. The treatment of this type of IR divergence is discussed in \cite{Anand:2020gnn}. A complete basis of states can be made out of local operators that are products of fermions where each fermion has at least one derivative acting on it; \cite{Anand:2020gnn} refer to this as the ``Dirichlet'' basis.  Each such state vanishes like $\sim p_i$ as $p_i \rightarrow 0$.  

As long as $m_q$ is not too small, this Dirichlet basis is sufficient for practical calculations.  However, when $m_q/g \ll 1$, although this basis is complete, it experiences extremely slow convergence with $\Delta_{\rm max}$.  The reason for the slow convergence is that the true eigenstates of the theory vanish at $p_i\rightarrow 0$ like $p_i^\alpha$ for some small value of $\alpha$.  Accurately approximating this nonanalytic behavior near the boundary requires a very large number of basis states that individually behave analytically. In the next subsection, we review this phenomenon at infinite $N_c$ as a warm-up to our treatment at finite $N_c$.

\subsection{Infinite $N_c$ Warm-Up}

Consider QCD with massive quarks in the large-$\Nc$ limit. We normalize the gauge coupling to one, $g^2 N_c / \pi = 1$, so the theory depends only on the parameter $m_q$. 
In this limit, the particle-number-changing interaction is suppressed by $\frac{1}{\Nc}$,
allowing us to simplify our basis by only including 2-particle states. We can
take the truncation $\Dmax$ high enough to quantitatively test the convergence
and illustrate all of the subtleties involved with nonzero mass. We will see 
that the LCT setup with the Dirichlet basis has good convergence if the quark mass
is not too small $(m_q \gtrsim 0.2 g\sqrt{\Nc/\pi})$. For small fermion mass, however, the 
boundary condition imposed by the Dirichlet basis leads to slow convergence, because it is trying to approximate $p^\alpha$ with a polynomial in $p$.

\begin{figure}[htbp]
\centering
\includegraphics[width=.5\linewidth]{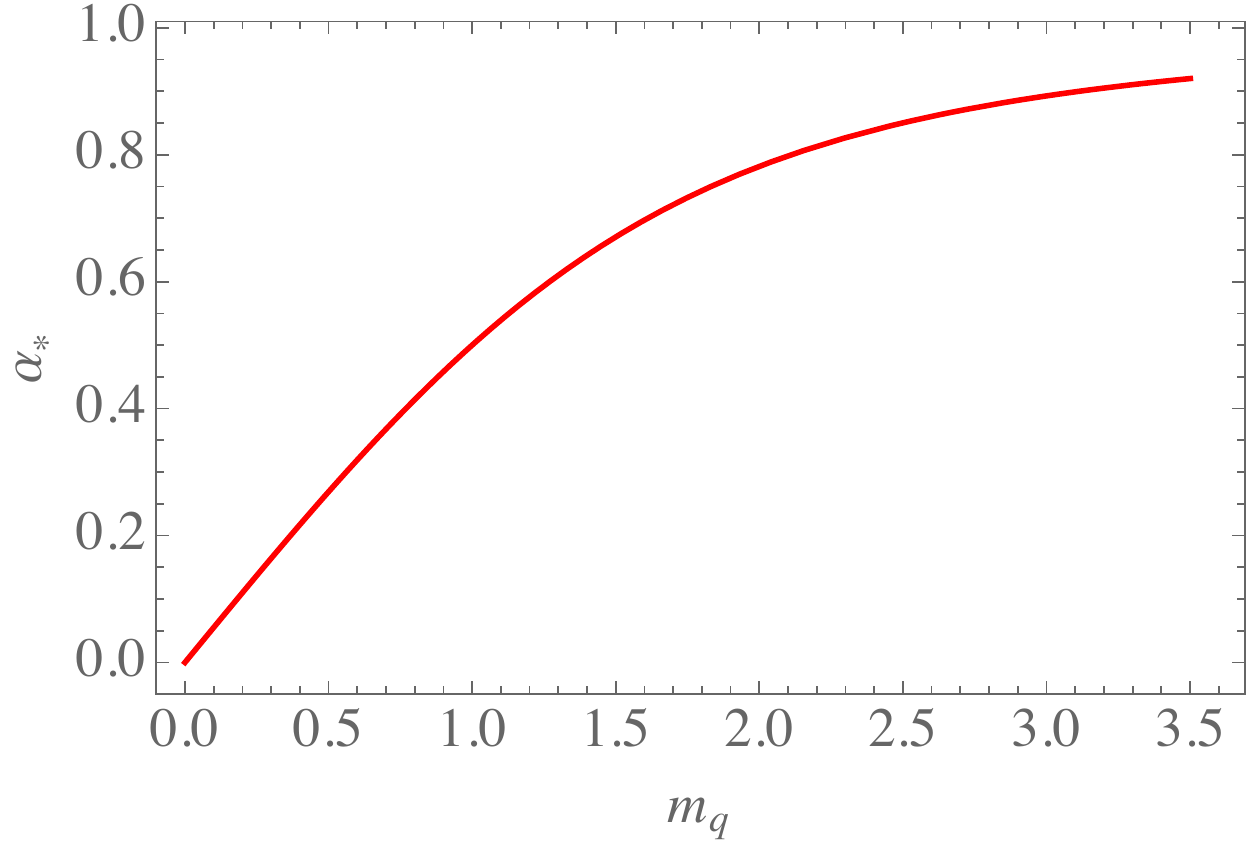}
\caption{\label{fig:plotHooftBC} The numerical values of $\alpha_*$ in the boundary asymptotic behavior $f(p_i, \cdots)\sim p_i^{\alpha_*}$, for the large $\Nc$ QCD states, according to (\ref{eq:HooftBC}).  
}
\end{figure}

The boundary condition of the large $\Nc$ QCD momentum space wave function was solved analytically in \cite{t1993two}. In the limit that either parton's momentum approaches zero, the wave function is a power law $p_i^{\alpha_*}$, where the power $\alpha_*$ is determined by the quark mass through
\begin{align}
\label{eq:HooftBC}
 \pi  \alpha_*  \cot \left(\pi  \alpha_*\right)+m_q^2=1 \, .
\end{align}
The Ansatz is then chosen to satisfy this boundary condition. The numerical values of $\alpha_*$ as a function of $m_q$ is plotted in Figure \ref{fig:plotHooftBC}. 
The true boundary condition smoothly interpolates between zero and $\CO(1)$ values. In comparison, the LCT non-Dirichlet basis has $\alpha=0$ and the Dirichlet basis has $\alpha=1$. 

In principle, one can choose any value of $\alpha$  with $0 \le \alpha \le 1$ and then define a set of LCT basis states that vanish like $p_i^\alpha$ at $p_i \sim 0$.  While all values of $\alpha>0$ will satisfy the Dirichlet boundary condition $f(0,\cdots)=0$ and thus work in principle, we expect only the ones close to `t Hooft's solution (\ref{eq:HooftBC}) will give the variational Ansatz a fast convergence rate. 

\begin{figure}[htbp]
\centering
\includegraphics[width=.6\linewidth]{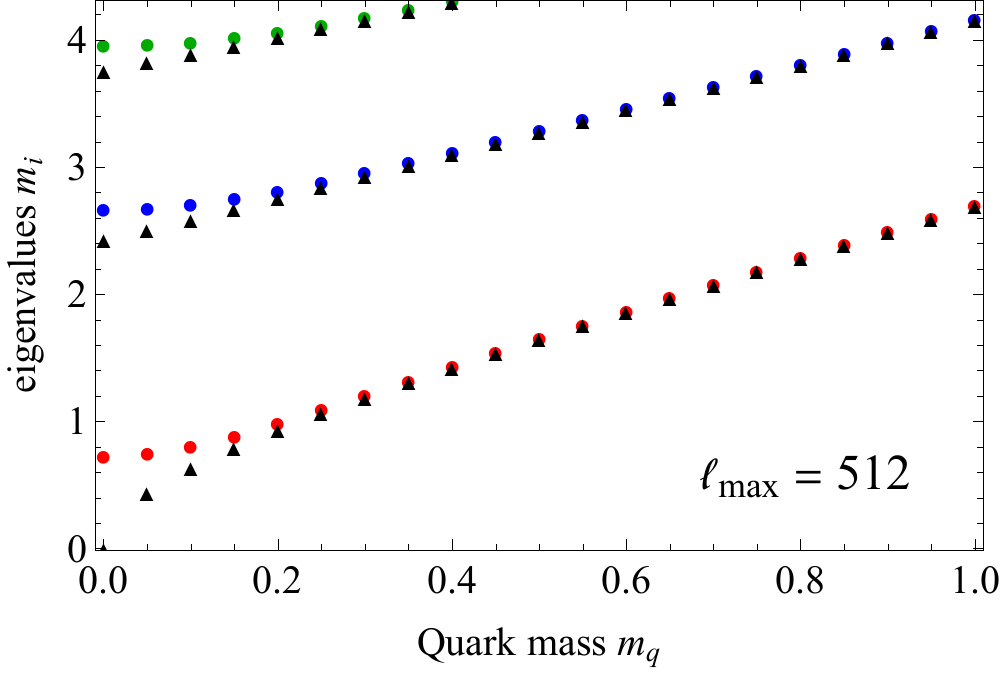}
\caption{\label{fig:massiveLargeNResult}
Low meson mass eigenvalues $m_i$ of the massive quark QCD in the large-$\Nc$ limit, as a
function of the quark mass $m_q$. The points are the LCT results computed with the 
Dirichlet basis  at $\ell_{\rm max}=512$. We use colors
red, blue and green for the lowest, second lowest and third lowest eigenvalues, respectively.
The black triangles are the known results using large $\Nc$ wave functions with `t Hooft's boundary condition. 
}
\end{figure}

As a particularly illustrative example of this slow convergence rate for the ``wrong'' choice of $\alpha$, consider the case where we use the Dirichlet basis (i.e. $\alpha =1$).
The Dirichlet basis states and their matrix elements are described in detail in \cite{Anand:2020gnn}.
At each value  of $m_q$  (in units where $g^2 N_c / \pi \equiv 1$), we diagonalize the Hamiltonian and obtain the mass eigenvalues.  Since LCT is a variational ansatz, we expect the low-lying spectrum to converge to the physical values at sufficiently high $\ell_{\max}$. We then plot these low-lying eigenvalues as a function and compare them against the known data computed from `t Hooft's ansatz. The result is shown in \ref{fig:massiveLargeNResult}. 

\begin{figure}[ht]
\centering 
\includegraphics[width=.48\linewidth]{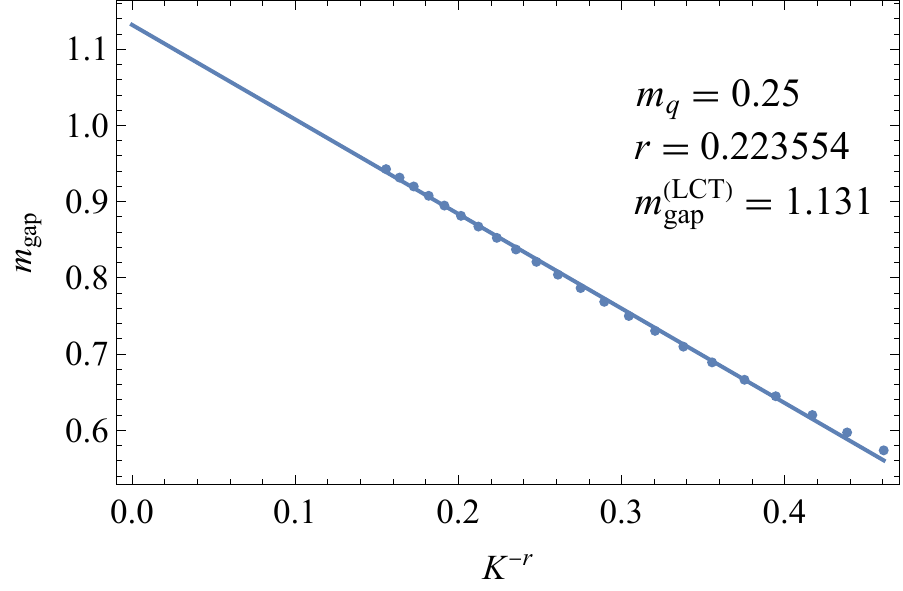}
\includegraphics[width=.48\linewidth]{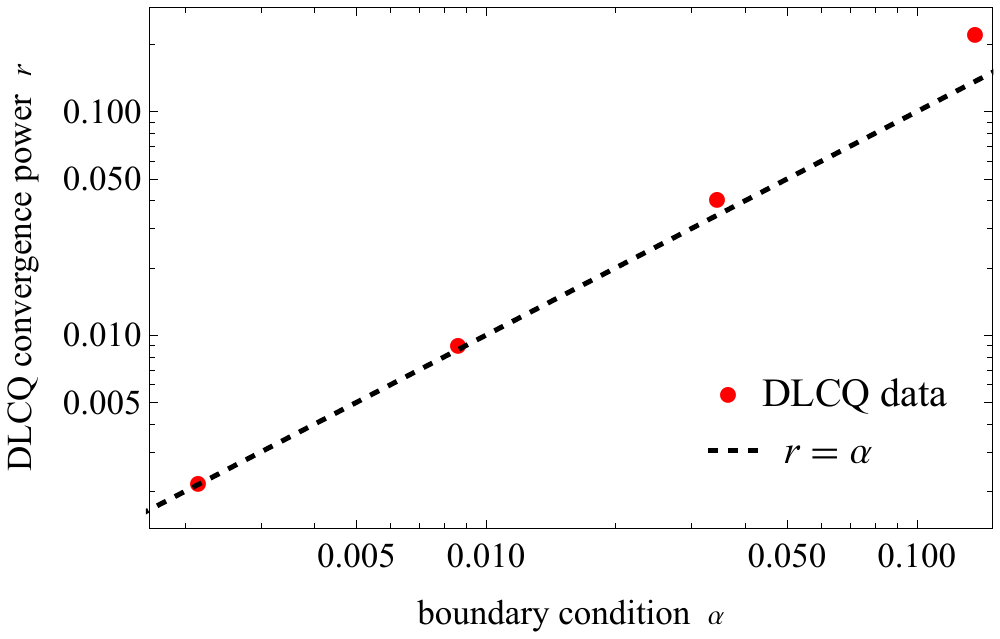}
\caption{\label{fig:DLCQ}
Measurements of the convergence DLCQ at large $\Nc$ at small $m_q$. 
{\bf Left: } An example of DLCQ convergence at $m_q=0.25$. The $y$-axis is the mass gap. The $x$-axis is the truncation $K$ to power $-r$. We use a fit $m_{\rm gap}^{\rm (DLCQ)}(K) = c K^{-r} + m_{\rm gap}(\infty)$ and take $m_{\rm gap}(\infty) \approx m_{\rm gap}^{(\rm LCT)}(\infty)$ from the LCT extrapolation described by Figure
\ref{fig:plotConvergenceBC}.
{\bf Right: } Compare $r$ in the the DLCQ convergence power law $K^{-r}$ with the boundary condition $\alpha$ computed from (\ref{eq:HooftBC}). The DLCQ data comes from repeating the procedures in the left plot to various $m_q$ and obtain the power law fit. A simple function $r=\alpha$ (black dashed line) fits the relation between $r$ and $\alpha$ quite well.
}
\end{figure}

The result shows that
the LCT result agrees well with the known data (black triangles) for $m_q \gtrsim 0.2$.
However,  in the small $m_q$ regime the LCT result starts deviating from the known result, and its lowest eigenvalue is nonzero at $m_q = 0$ even for the large truncation value  $\ell_{\rm max}=512$.  We expect that the error in the lowest eigenvalue should approach zero with increasing $\ell_{\rm max}$ at a rate $\sim \ell_{\rm max}^{-\alpha_*}$, so that as $m_q$ decreases, the convergence rate becomes worse and worse and eventually at $m_q=0$, where $\alpha_*=0$, the result using the Dirichlet basis never converges to the right answer. In fact, we see similar slow convergence in DLCQ with the truncation parameter $K$. In Fig.~\ref{fig:DLCQ}, we explicitly show the convergence rate of the lowest eigenvalue to its asymptotic value as a function of $K$.  For a range of quark masses, the convergence rate can be seen to be $K^{-\alpha_*}$,  as expected based on the behavior of the true lightest state wavefunction.

Our strategy will be to introduce nonlocal operators designed to behave like $p_i^{\alpha}$ at small $p_i$. In the new basis that we call {\it modified boundary condition basis}, the building blocks are the generalized free fields $\d^\alpha \psi$.  Their correlators are defined in terms of the following two-point function: 
\begin{equation}
\< \d^\alpha \psi (x) \d^\alpha \psi(y) \> = \frac{\Gamma(2\alpha+1)}{(x-y)^{2\alpha+1}} \, .
\end{equation}
Higher $n$-point functions are defined in terms of the two-point function through Wick contractions. We relegate the details of how to efficiently compute all the required Fourier transforms of correlators to Appendix \ref{app:ModifiedBasis}.

\begin{figure}[htbp]
\centering
\includegraphics[width=.6\linewidth]{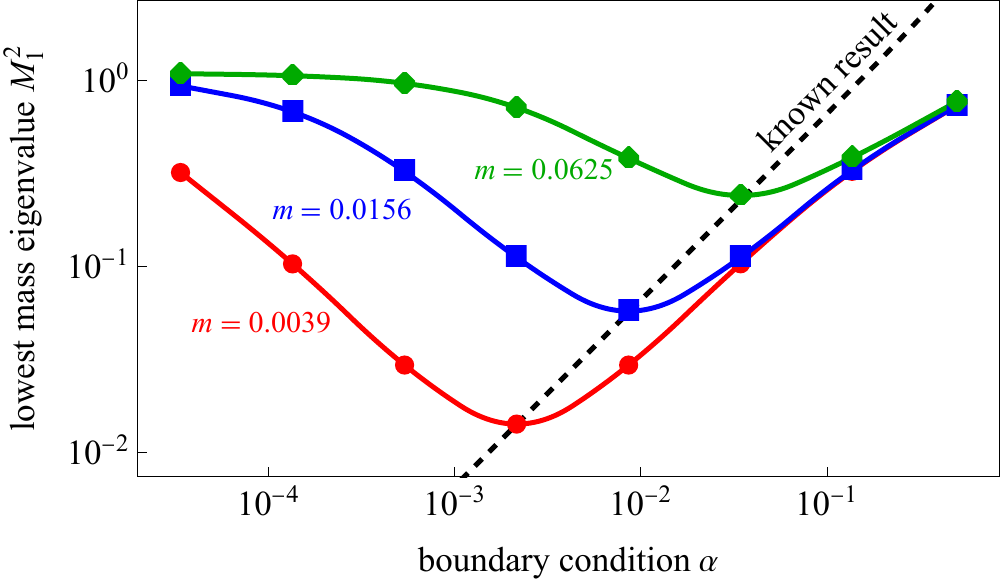}
\caption{
\label{fig:plotVariationBC}
Plot of the lowest eigenvalue, $M_1^{2}$ as a function of the boundary condition of the LCT generalized free field basis, $\alpha$. The red, blue, green lines represent results computed using three different quark masses $m_q$, respectively. The black dashed line is the known result given by the parametric line $(\alpha(m_q),M_1^2(m_q))$ where $\alpha(m_q)$ is `t Hooft's boundary condition (\ref{eq:HooftBC}) and $M_1^2(m_q)$ is the correct lowest eigenvalue. 
The known result line intersects each LCT line at the correct boundary condition with respect to its quark mass.
}
\end{figure}

\begin{figure}[htbp]
\centering
\includegraphics[width=.8\linewidth]{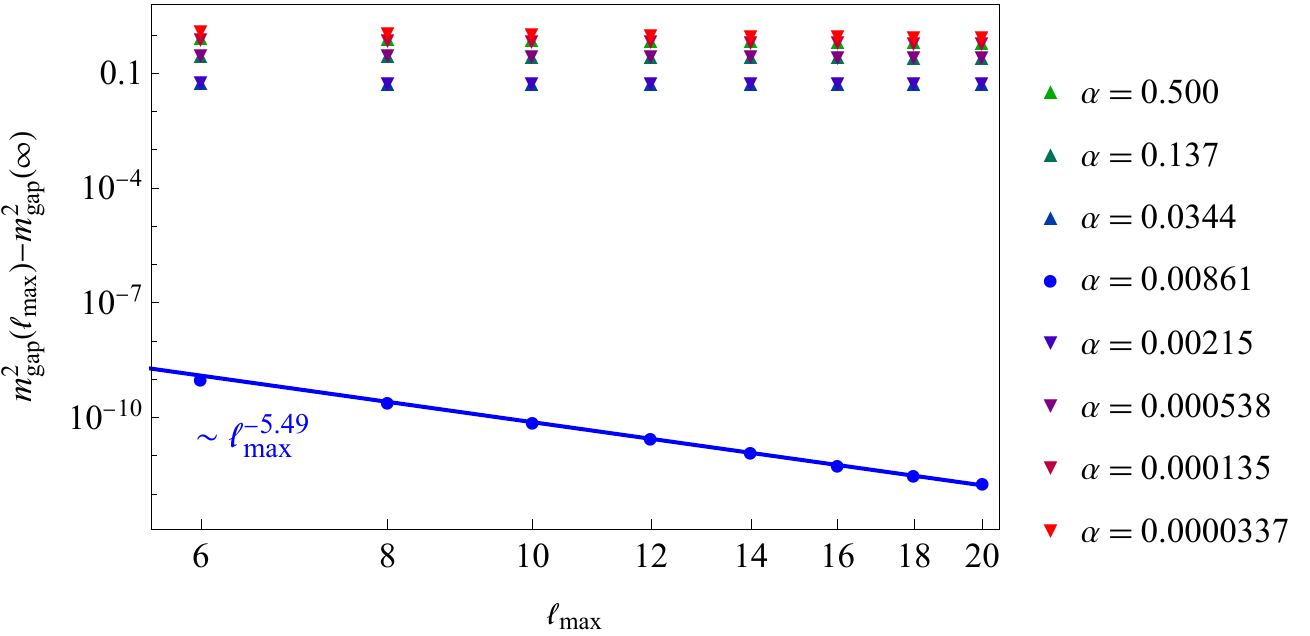}
\caption{
\label{fig:plotConvergenceBC}
Compare the convergence of the mass gap using different boundary condition $\alpha$. We take $m_q=0.015625$ and vary the boundary condition $\alpha$. 
For each $\alpha$, we diagonalize the Hamiltonian and obtain the mass gap $m_{\rm gap}$ and fit it as a function of $\ell_{\rm max}$. The fit model is a power law $m_{\rm gap}^2(\ell_{\rm max}) = m_{\rm gap}^2(\infty) + c_1 \ell_{\rm max}^{-c_2}$, with parameters $m_{\rm gap}^2(\infty)$, $c_1$ and $c_2$ to be determined by the fit. The boundary conditions $\alpha$ have a drastic effect on the convergence rate. The optimal boundary condition $\alpha_* = 0.00861$ (blue line), which agrees with (\ref{eq:HooftBC}). There the convergence is very fast, $\sim \ell_{\rm max}^{-5.5}$. Away from $\alpha_*$ (points of other colors), the convergence is poor.
}
\end{figure}

We test the basis by computing the large $\Nc$ QCD spectrum at different quark mass $m_q$, and for each $m_q$ we compute the truncated Hamiltonian using modified boundary condition basis at several different $\alpha$ values around the theoretical value $\alpha_*$. We then compare these different results with the theoretical values of from `t Hooft's boundary condition. The result is shown in Figure \ref{fig:plotVariationBC}. The result verifies that the convergence improves as the boundary condition $\alpha$ of the basis approaches $\alpha_*$, and that the modified boundary condition LCT result matches the theoretical result. We also show the convergence of the lowest mass eigenvalue as a function of the truncation order $\ell_{\rm max}$, taking different $\alpha_*$, for some small $m_q$. This is shown in Figure \ref{fig:plotConvergenceBC}. When using the correct boundary condition, the LCT method has a convergence rate of $\ell_{\rm max}^{-5.5}$, which is significantly better than the Dirichlet basis.

\section{Finite $N_c$}
\label{finiteNbasis}

In the last section we saw that the LCT method successfully reproduces the large $\Nc$ known results. In this section we will use the LCT method to study finite $\Nc$ QCD.   With $\Nc$ finite, particle number-changing matrix elements are no longer suppressed by large $\Nc$ and we must include multiparticle states.

\subsection{Boundary Condition Implementation}

As is discussed in the previous section, the fermion wave function has a power law boundary condition as the individual parton momentum fraction $p_i/p_{\rm tot} \rightarrow 0$, and at small quark mass the convergence of truncation results is good only if we choose an ansatz with the correct boundary condition. At finite $\Nc$, the number of states grows rapidly with the increasing particle number, so we need to compute the basis and matrix elements efficiently.

We implement the full-fledged LCT method with a generalized free field basis. In the process of building the basis, we treat $\d^\alpha\psi$ as a primary operator, where $\alpha$ is a real number which we will determine later using adaptive methods. The full basis is constructed recursively by sewing two lower dimension primary operators $A$ and $B$ to form a double-trace operator 
\begin{equation}
\left[ AB \right]_\ell \equiv \sum_{m=0}^\ell c^\ell_m(\Delta_A,\Delta_B)\, \p^m A \, \p^{\ell-m} B,
\end{equation}
where the coefficients $c^\ell_m(\Delta_A,\Delta_B)$ are given by the formula
\begin{equation}
c^\ell_m(\Delta_A,\Delta_B) = \fr{(-1)^m \Gamma(2\Delta_A+\ell) \Gamma(2\Delta_B+\ell)}{m! (\ell-m)! \Gamma(2\Delta_A+m) \Gamma(2\Delta_B + \ell - m)}  \,.
\end{equation}
We find selection rules to optimize the efficiency of constructing the basis in the charge-neutral sector. We begin with bosonic operators $\phi_n \equiv \left[\d^\alpha\psi^\dagger \d^\alpha\psi\right]_n$. Each $\phi_n$ spans a basis of multi-boson primaries similarly to that of a free scalar, and the full basis is spanned by taking arbitrary multi-trace operators between primaries in different $\phi_n$ bases. At large $\Nc$ this basis is the full minimal basis since all $\phi_n$ are truly independent. Finite $\Nc$ reduces the space a bit more, and we need to compute the Gram matrix to eliminate the redundancy due to algebraic constraints. Because primaries at different dimensions are automatically orthogonal, we only need to compute the Gram matrix within each dimension sectors. When we compute the matrix elements, we decompose each primary operator into a linear combination of ``monomials''
\begin{equation}
\CO \equiv \sum_{\kvec} C_{\kvec}^{\CO} (\d^{k_1+\alpha}\psi^\dagger\d^{k_2+\alpha}\psi)\cdots
(\d^{k_{2n-1}+\alpha}\psi^\dagger\d^{k_{2n}+\alpha}\psi),
\end{equation}
and we need to compute the matrix elements between every pair of monomials. 
The space of monomials is more redundant than that of the primary operators, because they also span all descendant states. In  momentum space, the descendant states are not linearly independent from the primary states, which means we can use the Gram matrix to eliminate the redundant monomials. 
Finally, we use the Wick contraction to compute the matrix elements. The Wick contraction is much more efficient than the Fock space integrals as it reduces the integrals between every pair of partons to
one expensive principal value integral of only one pair of variables $dx$ and $dx'$ and leaves the spectator partons as a simple algebraic expression that can be calculated recursively. 

\subsection{Determining the Correct Boundary Condition}

\begin{figure}[ht]
\centering
\includegraphics[width=0.6\linewidth]{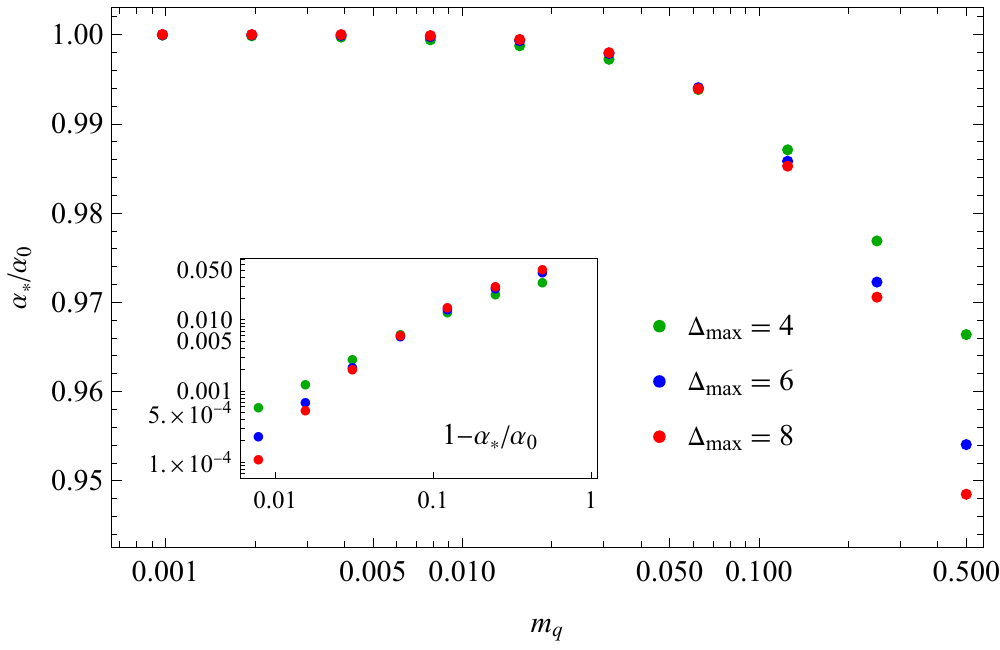}
\caption{
    \label{fig:variationalBC}
    A plot of the optimal boundary condition $\alpha_*$ determined by the variatioal method at different truncation $\Dmax$, normalized by the initial approximation $\alpha_0$. At small $m_q$, $\alpha_*$ has already perfectly converged at small $\Dmax$.
}
\end{figure}

\begin{figure}
\centering
\includegraphics[width=0.48\linewidth]{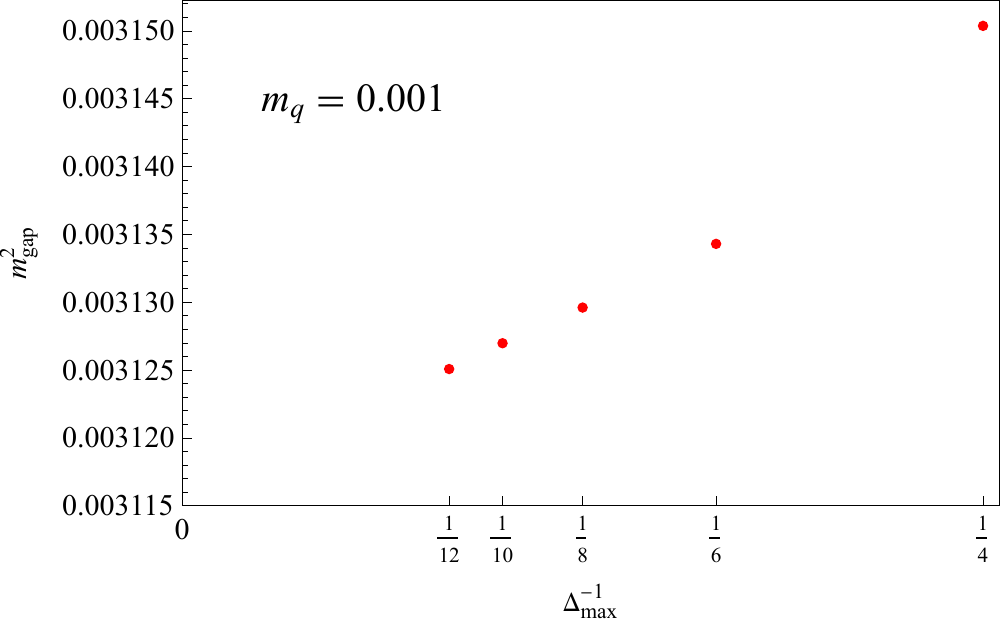}
\includegraphics[width=0.48\linewidth]{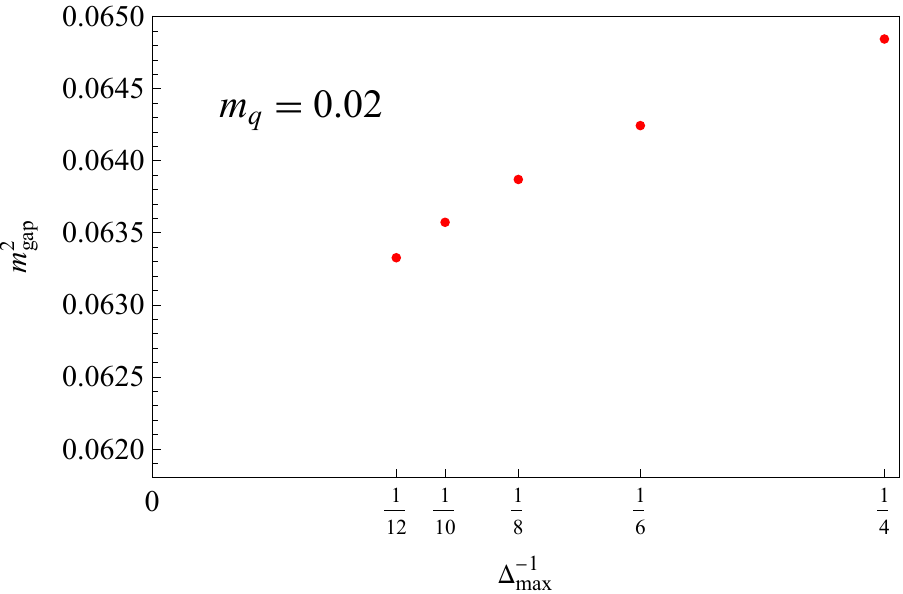}
\includegraphics[width=0.48\linewidth]{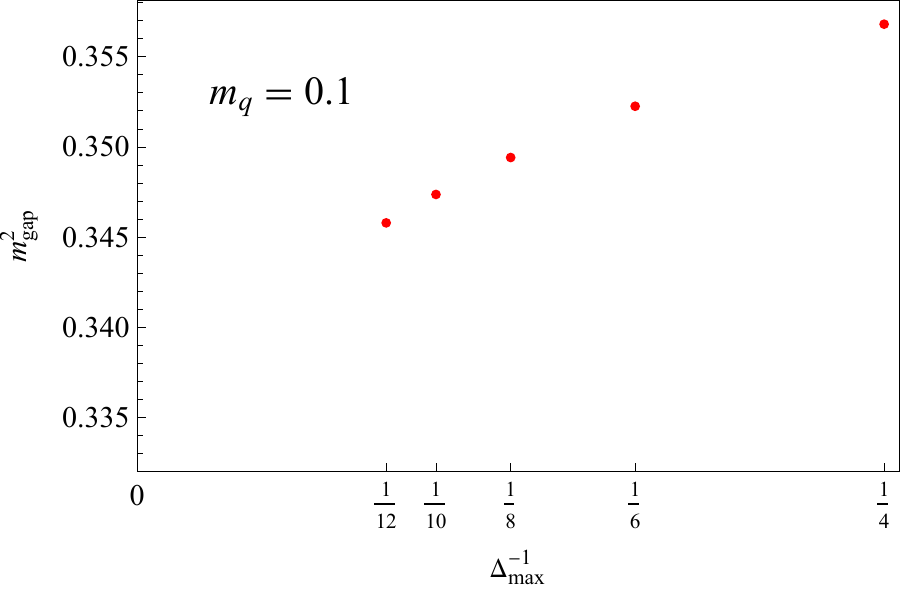}
\includegraphics[width=0.48\linewidth]{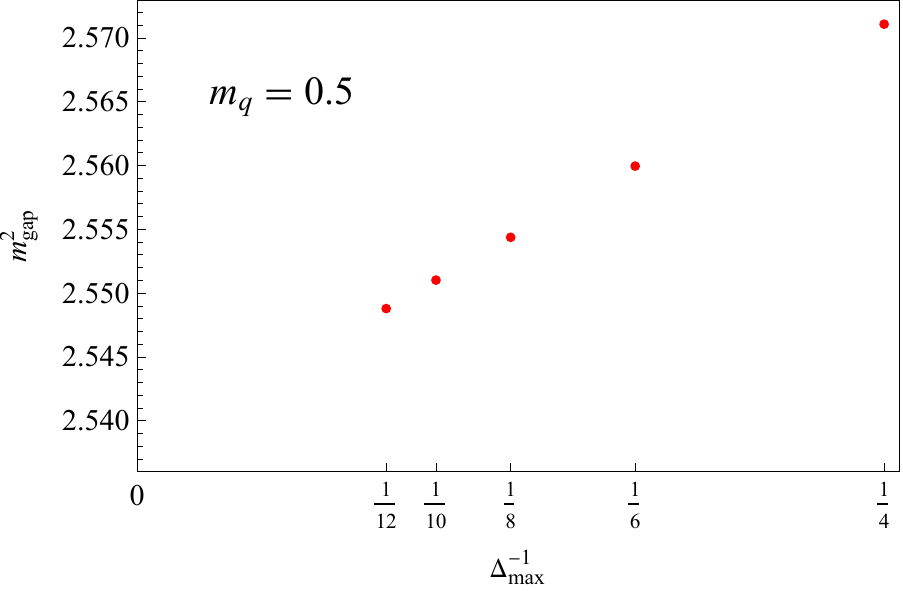}
\caption{
    \label{fig:gapConvergence}
    A plot of the convergence of the mass gap at different $m_q$. In all four cases, the mass gap has converged to percent level accuracy, and even better for $m_q=0.001$. 
}
\end{figure}

We will use a variation method to determine the correct boundary condition $\alpha$ numerically. Hamiltonian Truncation is a variational ansatz, hence the ground state eigenvalue is always overestimated by finite truncation. We will take $\alpha$ as the variational parameter, and change $\alpha$ adaptively until we find the value $\alpha_*$ that minimizes the mass gap. 
An approximation $\alpha_* \approx \alpha_0 = \frac{\sqrt{3}m_q}{\sqrt{1+\Nc^{-2}}\pi} + \CO(m_q^2)$ 
can be found \cite{Sugihara} by restricting to two-particles, and solving the Bethe-Salpeter equation with finite $\Nc$ at small parton-$x$.\footnote{More generally, the full expression for $\alpha_0$ is given by the solution to the equation $\pi \alpha_0 \cot ( \pi \alpha_0) + \frac{m_q^2}{1- N_c^{-2}} = 1$, see \cite{Sugihara}. We emphasize that $\alpha_0$ is defined as the boundary condition in the two-particle truncation, whereas $\alpha_*$ is defined as the true boundary condition in the full theory.} We will take $\alpha_0$ as the initial seed of the search. The result is shown in Figure~\ref{fig:variationalBC}. The good news is that $\alpha_*$ converges fast enough so that we can perform the expensive variational search at low $\Dmax$ and use the same $\alpha_*$ to compute the matrix elements at high $\Dmax$.  Once we have chosen an optimal (or nearly optimal) value of $\alpha$ for the construction of our basis, the gap converges reasonably ($\sim \Delta_{\rm max}^{-1}$) as a function of $\Delta_{\rm max}$, as we show in Fig.~\ref{fig:gapConvergence}.

\section{Comparison of Truncation Results with Chiral Lagrangian}
\label{results}

Now we have addressed all the conceptual and technical issues in studying massive 2D QCD at finite $\Nc$ using LCT. In this section we present these numerical results and, as promised, compare the truncation data at small $m_q$ with the Chiral Lagrangian prediction. 
In section \ref{sec_SG} we argued that the Chiral Lagrangian describes the Sine-Gordon model (\ref{eq:SG-Lagrangian}), which means we know the spectrum exactly at small $m_q$. 

\begin{figure}[htbp]
\centering
\includegraphics[width=0.48\linewidth]{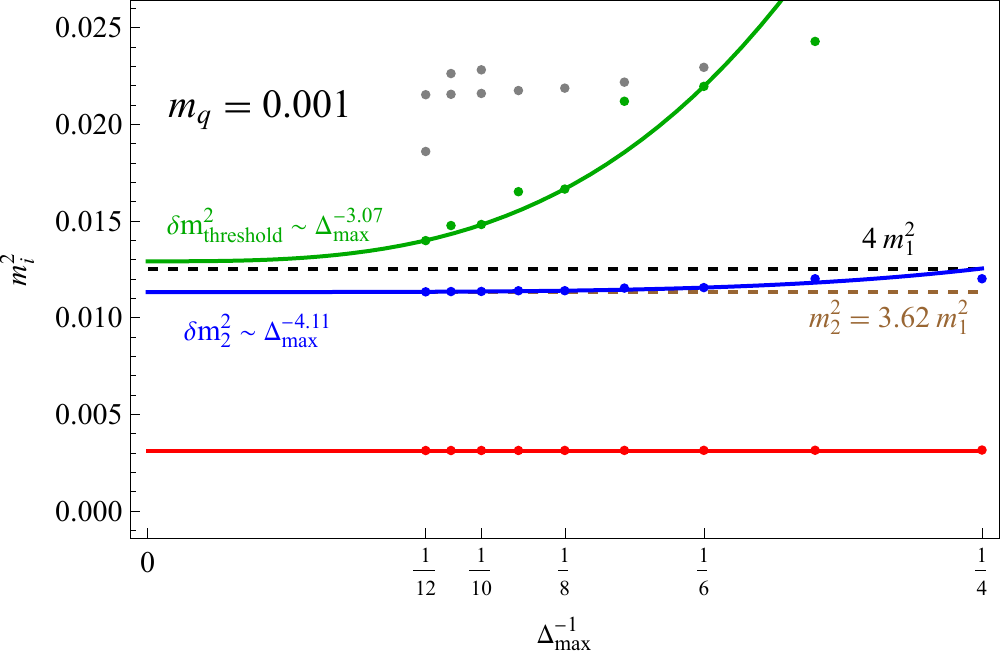}
\includegraphics[width=0.48\linewidth]{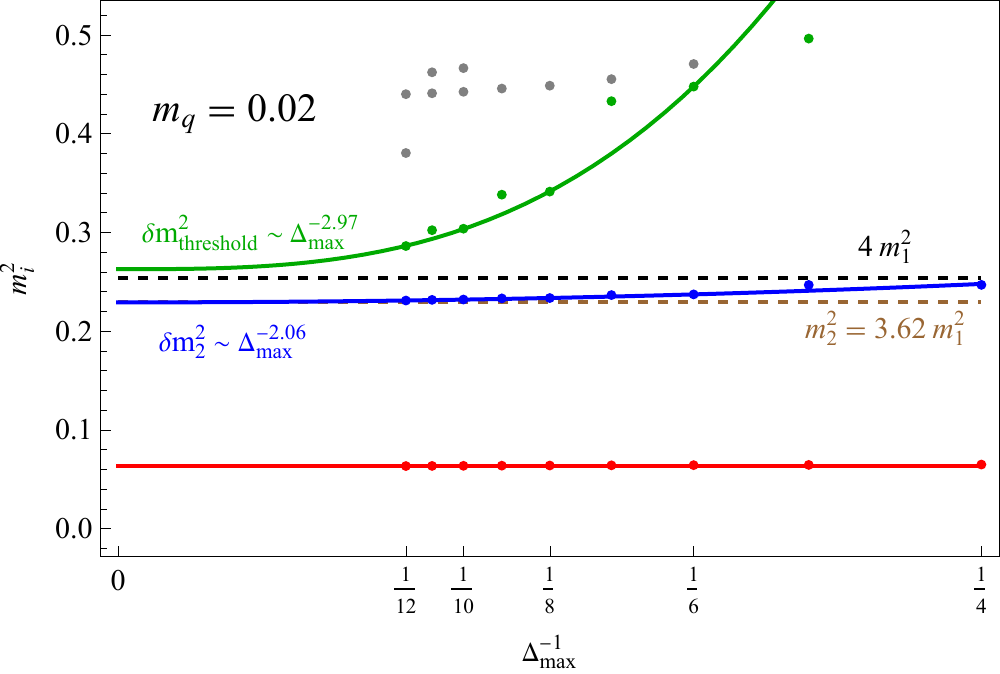}\\ ~\\
\includegraphics[width=0.48\linewidth]{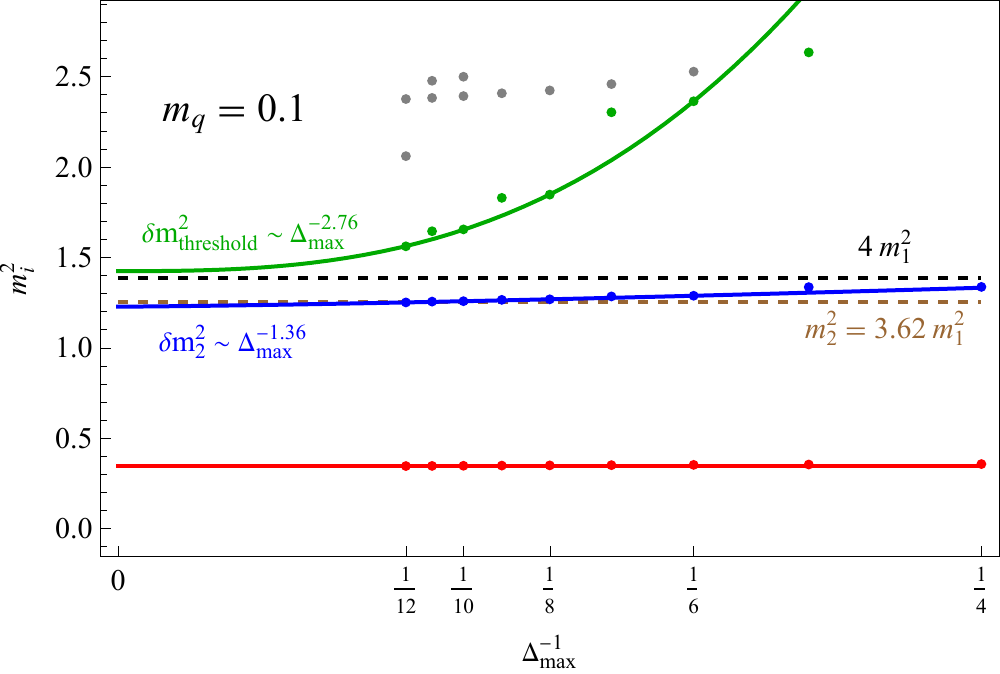}
\includegraphics[width=0.48\linewidth]{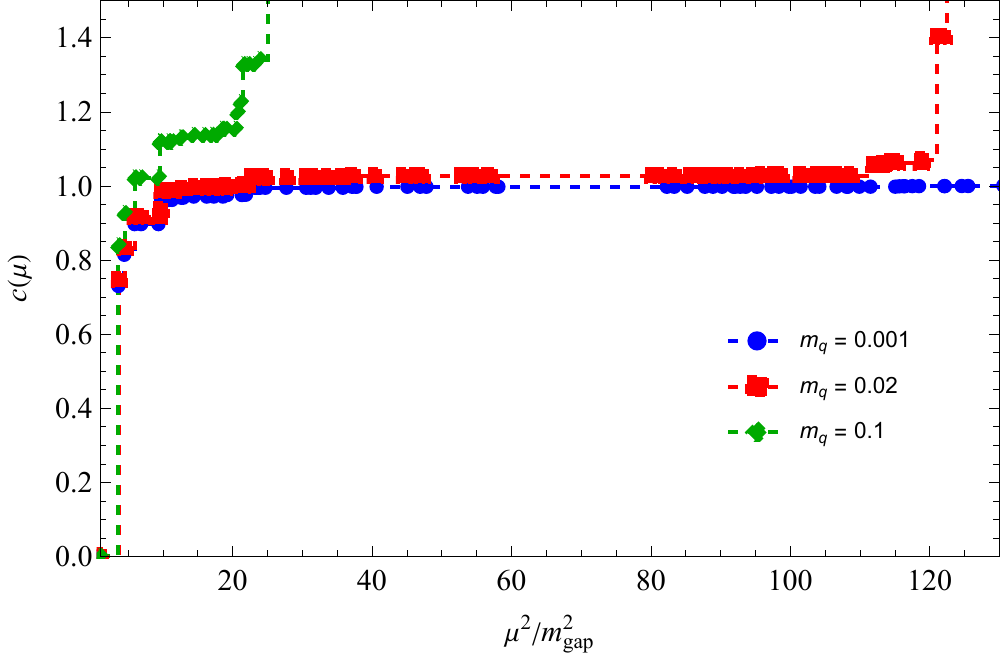}
\caption{
    \label{fig:massiveDataCollection}
    Spectrum and c-function of QCD at $\Nc=3$ for different quark masses $m_q$. 
    {\bf Upper left: } The mass eigenvalues as a function of the truncation $\Dmax$, at $m_q=0.001$. The lowest three eigenvalues are highlighted with red, blue and green points, respectively. 
    We fit the blue and green data at even $\Dmax$ to $m^2 = c_1 + c_2 \Dmax^{-c_3}$ and show the fit as blue and green lines, respectively. The red line is $m_{\rm gap}(\Dmax = 12)$, and its convergence is shown separately in Figure \ref{fig:gapConvergence}.
    These trend lines predict the eigenvalues in the $\Dmax\rightarrow \infty$ limit. 
    The gray points are higher eigenvalues. The black dashed line is the two particle threshold, and the brown dashed line is the Sine-Gordon theoretical prediction of the second bound state. The second eigenvalue (blue line) clearly deviates from the two particle threshold (black dashed line) and converges to the Sine-Gordon prediction (brown dashed line) to a good approximation.
    {\bf Upper right: } The mass eigenvalues as a function of the truncation $\Dmax$, at a larger quark mass $m_q=0.02$.
    {\bf Lower left: } The mass eigenvalues as a function of the truncation $\Dmax$, at a larger quark mass $m_q=0.1$.
    {\bf Lower right: } The c-function (y-axis) of the three quark masses shown in the same plot, using the mass gap as the unit of the energy scale (x-axis). The blue, red and green lines correspond to $m_q = 0.001, 0.02$ and $0.1$, respectively. At small $m_q$ the c-function has a clear plateau at $c=1$, and the UV correction sets in at lower scale for increased $m_q$.
}
\end{figure}

At $\Nc=3$, we expect 5 bound states, two of which have mass below the two particle threshold. Figure \ref{fig:massiveDataCollection} shows the spectrum and the c-function of three different choices of $m_q$. For the $m_q=0.001$ and $m_q=0.02$ spectrum plots, the three highlighted eigenvalues (red, blue, green, respectively) converge to the Sine-Gordon prediction values to a good accuracy, while the $m_q=0.1$ the same eigenvalues have a noticeable deviation. This observation is consistent with the c-function plot. Both $m_q=0.001$ and $m_q=0.02$ c-functions have a clear plateau at $c=1$, which matches the central charge of the free massless scalar. The $m_q=0.02$ c-function has a small correction to the plateau central charge, and the UV states enter earlier than that of $m_q=0.001$. The $m_q=0.1$ c-function has almost no plateau. We conclude that $m_q=0.001$ is small enough to match to Sine-Gordon in the IR. By contrast, Sine-Gordon gets a small correction at $m_q=0.02$ and a large correction at $m_q=0.1$. 

\begin{table}[htbp]
\centering
$
\begin{array}{|c||c|c|c|c|c|}
\hline
 n_{\max} & 4 & 6 & 8 & \text{no limit} & \text{Sine Gordon} \\
\hline
 m_1 & 0.0033341 & 0.00312532 & 0.0031251 & 0.00312507 & - \\
\hline
 m_2 & 0.0126691 & 0.0115498 & 0.0113329 & 0.0113315 & - \\
\hline 
m_2/m_1 & 1.94932 & 1.92238 & 1.90431 & 1.90421 & 1.90211 \\
\hline
 E_{\rm bind}/m_1 & 0.05068 & 0.07762 & 0.09568 & 0.09579 & 0.09789 \\
\hline
\end{array}
$
\caption{
    \label{tab:bindingEnergyTruncateParticleNumber}
    Binding energy $E_{\rm bind} \equiv 2m_1-m_2$ of the second bound state measured from the LCT Hamiltonian restricted to $n_{\rm max}$ particles. We use truncation $\Dmax = 12$ and small quark mass $m_q=2^{-10}\approx0.001$. The LCT result accurately reproduces the Sine-Gordon theory prediction of the binding energy for $n_{\rm max} = 8$ or higher, while the measurement with particle number restricted to 4 has a significant error.  Our $\Delta_{\rm max}=12$ basis has 77 states in the light sector (i.e.~become massless as $m_q \rightarrow 0$) and 741 in the heavy sector.
}
\end{table}

Now we take a closer look at the small $m_q$ regime. We take a tiny quark mass, $m_q = 2^{-10} \approx 0.001$. At this $m_q$ the IR data is dominated by Sine-Gordon and the finite $m_q$ correction is negligible, so we are able to compare the LCT result and the Sine-Gordon result quantitatively.
In Table \ref{tab:bindingEnergyTruncateParticleNumber} we show the ratio between two lowest bound state masses $m_2/m_1$, and compare it with the ratio $m_2/m_M$ from the Sine-Gordon prediction (\ref{eq:SineGordonMassPrediction}). Our full-fledged truncation result at $\Delta_{\rm max}=12$ reproduces the Sine-Gordon binding energy to about $2\%$ accuracy.\footnote{At $\Delta_{\rm max}=4,6,8,10$ and $12$, the number of $(\textrm{light},\textrm{heavy})$ states is $(5,7), (11,29), (22,95), (42, 277),$ and $(77,741)$, respectively.} Because the binding energy itself is small -- about $5\%$ -- this level of accuracy in the binding energy required a significantly higher level of accuracy, of about $0.1 \%$, in the total energy of the bound state.  The inclusion of high-particle-number states was absolutely necessary for this result.  The prior work \cite{Sugihara} used a different truncation similar to LCT but where the number of particles was limited to 4 or less. As one can see from the above table, with at most 4 particles, the binding energy has a significant deviation from the Sine-Gordon prediction, by about a factor of 2.

\begin{figure}[htbp]
\centering
\includegraphics[width=0.6\linewidth]{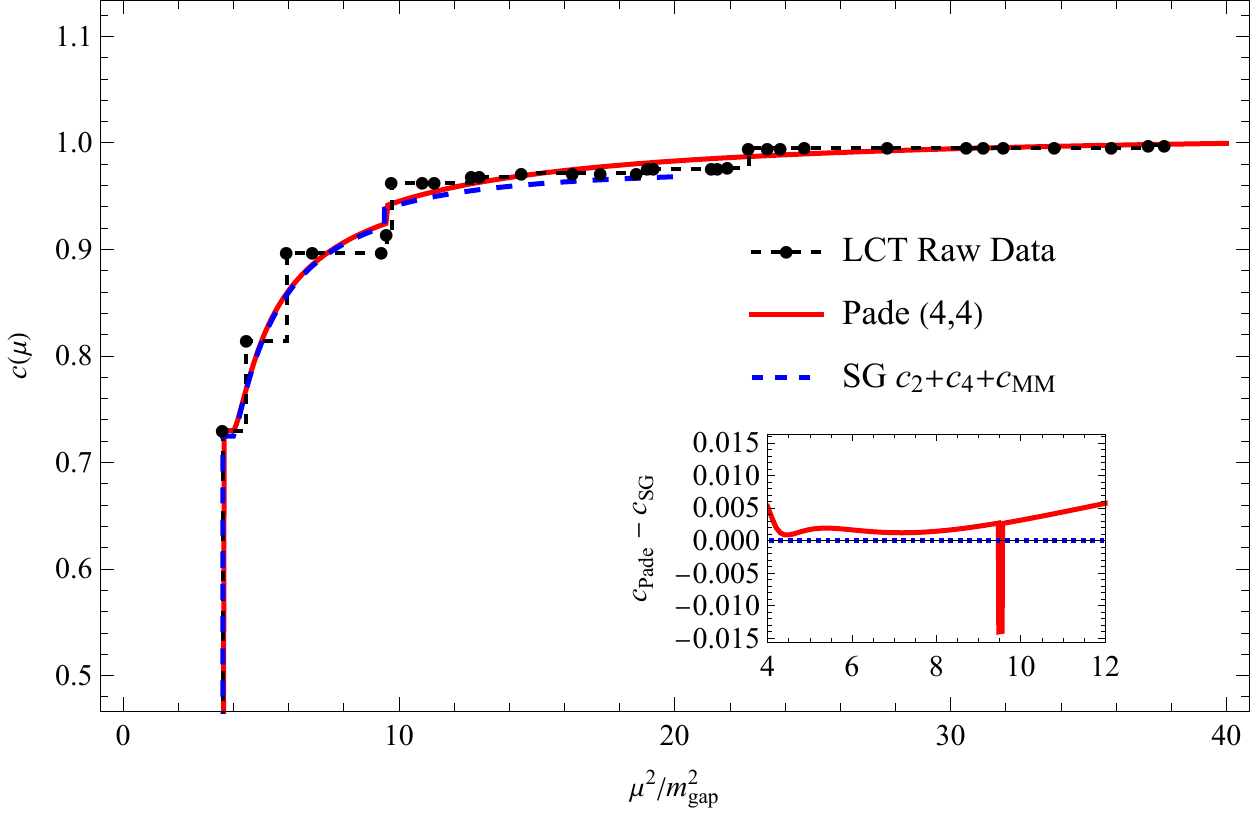}
\caption{
    \label{fig:pltCFuncSG} 
    The IR $c$-function at $m_q = 2^{-10}\approx0.001$. The discrete points connected by the dashed line is the LCT raw data computed using (\ref{eq:spectralDensityDefinition}). The red solid line is the interpolation using analyticity and $(4,4)$-Pad\'e approximant. The blue dashed line represents the Sine-Gordon integrability result including the contribution from the pion two particle states $c_{MM}$, and the one particle states of parity even breathers, $c_2$ and $c_4$. The $c_{MM}$ contribution is continuous, and $c_2$ and $c_4$ are the two kinks at $m_2^2=3.61 m_{\rm gap}^2$ and $m_4^2=9.47 m_{\rm gap}^2$; the small spike-like feature in the inset is due to a sub-percent level difference in $m_4^2$ between the truncation and Sine-Gordon calculation. The Sine-Gordon result is accurate up to the soliton-anti-soliton threshold $\mu^2=10.47 m_{\rm gap}^2$. Up to this energy, the LCT result follows the Sine-Gordon to a very good accuracy. Beyond the threshold, one should also add soliton-anti-soliton contributions to the Sine-Gordon $c$-function, which are not included in the plot above.
}
\end{figure}

For any $m_q$, we can compute the spectral density of the stress tensor, and use it to compute the Zamolodchikov $c$-function.
At small $m_q$, we can compare the c-function quantitatively between the Sine-Gordon prediction and the truncation results. On the Sine-Gordon side, the c-function can be constructed from the form factors of the stress tensor. The larges contribution comes from the meson-meson 2-particle continuum $c_{MM}$, and the parity even bound states $c_2$ and $c_4$.
The $c_{MM}$ contribution is (see e.g.~\cite{Karateev:2019ymz})
\begin{equation}
c_{MM}(s) = 12\pi \int_0^{\bar\theta} \frac{d\theta}{4\pi} 
\frac{1}{4\cosh ^4 \frac{\theta }{2} }
|F^{\Theta}_{b_1b_1}(\theta)|^2  \, , \quad \bar\theta\equiv 2 \,{\rm arccosh}\frac{\sqrt{s}}{2}
\end{equation}
where $F^{\Theta}_{b_1b_1}$ is the form factor of the trace of the stress tensor with the pion-pion two particle continuum
\begin{equation}
F^{\Theta}_{b_1b_1}(\theta) = 
\frac{2\cos\frac{\pi \nu}{2} }{\cos\pi\nu - \cosh\theta}
\cosh \left( \frac{i \pi-\theta}{2}  \right)
\exp \left(
    \int \frac{2}{t}
    \frac{\cosh \left(\nu -\frac{1}{2}\right) }{ \sinh t \cosh \frac{t}{2}}
    \sin ^2\frac{(i \pi-\theta ) t}{2 \pi }
\right) \, .
\end{equation}
The total contribution is $c_{MM}(\infty) = 0.235420$.
The contribution from the bound states are predicted using the results in \cite{Babujian:1998uw,Babujian:2001xn}
\begin{equation}
\label{eq:BoundStateFormFactors}
\begin{aligned}
&\frac{m_2^{(\rm SG)}}{m_{\rm gap}} = 1.90211, \quad \frac{m_2^{(\rm LCT)}}{m_{\rm gap}} = 1.90421, \quad c_2^{\rm(SG)} = 0.724596, \quad c_2^{\rm(LCT)} = 0.729811, \\
&\frac{m_4^{(\rm SG)}}{m_{\rm gap}} = 3.07768, \quad \frac{m_4^{(\rm LCT)}}{m_{\rm gap}} = 3.09220, \quad c_4^{\rm(SG)} = 0.016748, \quad c_4^{\rm(LCT)} = 0.016592 \, .
\end{aligned}
\end{equation}
See Appendix \ref{sec:FormFactors} for details. 
The combination of $c_{MM}$, $c_2$ and $c_4$ is accurate up to the soliton-anti-soliton threshold $s = \left( \frac{m_{\rm gap}}{ \sin\frac{\pi\nu}{2}} \right)^2 = 10.47 \times m_{\rm gap}^2$. For energy greater than the threshold, the above result is still a good approximation since they add up to $96\%$ of the central charge.
On the truncation side, we use (\ref{eq:spectralDensityDefinition}) to compute the spectral density of $T_{--}$.  The poles at $s=m_2^2$ and $m_4^2$ are identified as the eigenstates nearest to the predicted $m_i$ with non-zero contribution to the $c$-function. The rest of the states belong to the continuum. 
When $m_q$ is comparable with the UV scale $g \sqrt{\Nc/\pi}$, the IR physics gets a significant correction. The behavior of the spectral density is different from that of the Sine-Gordon model. Some of the Sine-Gordon bound states are no longer stable due to the interaction with UV degrees of freedom, and they form resonances in the multi-particle continuum. 

Because of the truncation of the Hilbert space, the spectrum of eigenvalues is discrete and a literal implementation of the spectral density (\ref{eq:spectralDensityDefinition})  produces a discrete approximation of the true $c$-function. We can interpolate the $c$-function by taking advantage of the analyticity of the time-ordered correlator \cite{Chen:2021bmm}. First, we use the discrete spectral density $\rho_{T_{--}}(\mu_i)$ to compute the time-ordered correlator 
\begin{equation}
G(s) = \sum_i \frac{\rho_{T_{--}}(\mu_i)}{s - \mu_i^2 + i\epsilon} .
\end{equation}
Away from physical poles and branch cuts, the time-ordered correlator is an analytic function of $s$, and its Taylor coefficients in $s$ around, for instance, $s=0$, converge as $\Delta_{\rm max} \rightarrow \infty$.  From the Taylor series and the analytic properties of the correlator, we can reconstruct its behavior in the full complex plane.  
In practice,  we take the coordinate transformation
\begin{equation}
s = 4 m_{\rm gap}^2 \frac{4 z}{(1+z)^2}
\end{equation}
so the multi-particle branch cut is mapped to the unit circle. Moreover, $G(s)$ is a sum of isolated physical poles, with truncation eigenvalues $\mu_i < 2 m_{\rm gap}$, plus a function of $z$ that is analytic in the unit disk.  The basic idea is to approximate $G(s)$ (minus the stable particle poles) by a Pad\'e approximant in the variable $z$, denoted as $\tilde G(s)$. 
Then, we map back to $s$-coordinate and take the imaginary part $\Im \tilde G(s)$ to obtain the interpolation of our spectral density
\begin{equation}
\tilde \rho_{T_{--}}(s) = -\frac{1}{\pi} \Im \tilde G(s) \, .
\end{equation}
We show the comparison between the LCT result and the Sine-Gordon prediction for the $c$-function in Figure \ref{fig:pltCFuncSG}. 

\begin{figure}[t]
\begin{center}
\includegraphics[width=0.9\textwidth]{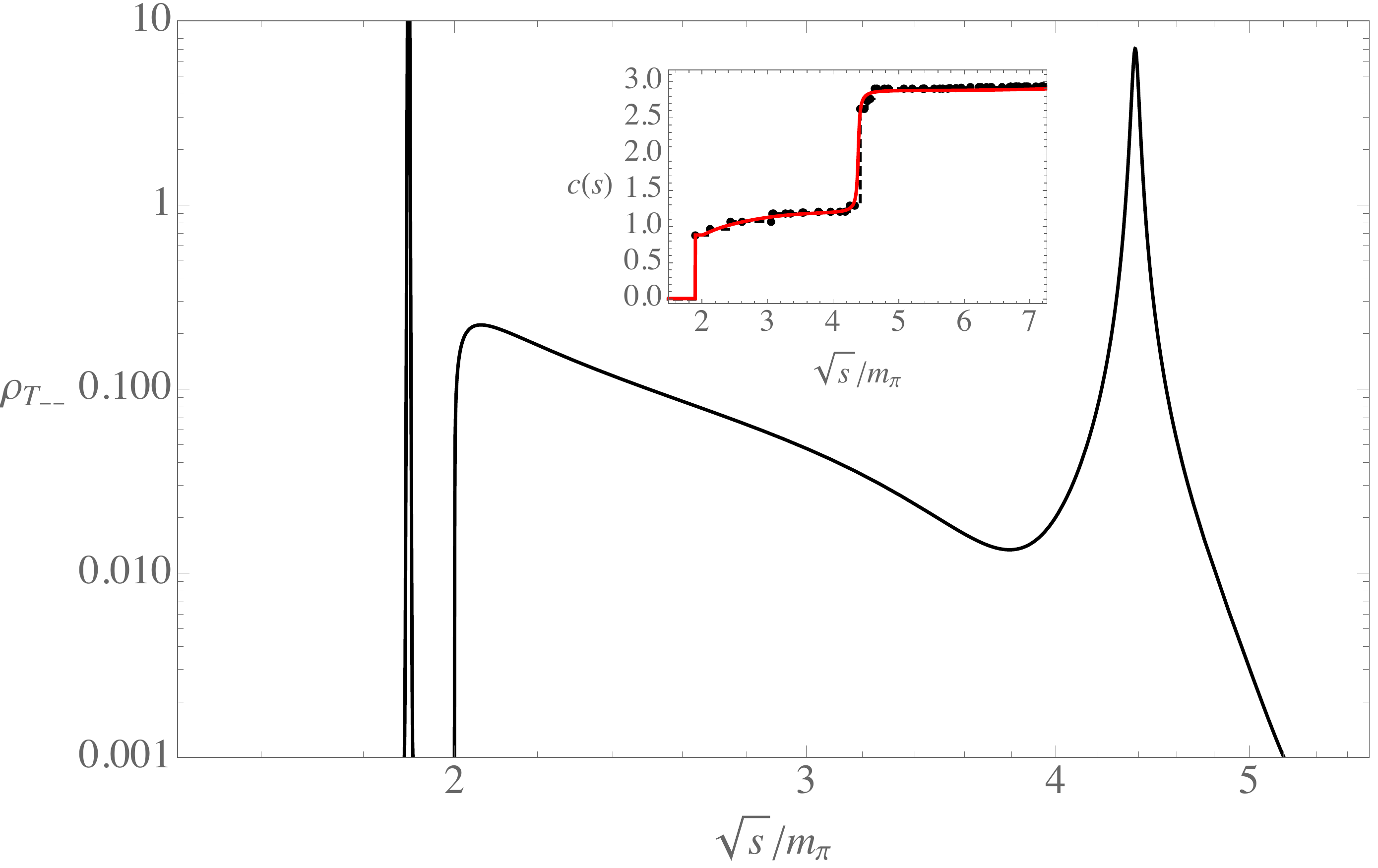}
\caption{Spectral density for $T_{--}$ from truncation, at $m_q = 0.125$ and $N_c=3$.  The bound state $\delta$ function has been given a very small width to make it visible to the eye.  The multi-particle continuum and resonance peak require processing the discrete spectral results from truncation as described in the text. In the inset, we show the comparison of the integrated spectral density, which in this case is the $C$-function, from the raw LCT data ({\it black dots}) and from integrating the processed spectral density ({\it red, solid}).}
\label{fig:pltSDSG}
\end{center}
\end{figure}

At larger values of $m_q$, there are unstable particles that appear as resonances in the spectral density.  In this case, it is important to improve the performance of the Pad\'e approximant by first subtracting out bound states and resonances from $G(s)$, and then taking a Pad\'e approximant of the remainder.\footnote{The function $G(s)$ minus the stable particle poles is still an analytic function of $z$ in the unit disk even when there are complex poles from unstable resonances. The decay width of the resonances pushes their poles down in the complex plane through the multiparticle branch cut onto the second sheet in $s$, which is outside the unit disk in $z$.  So, in principle with high enough $\Delta_{\rm max}$, the Pad\'e approximation would automatically capture the resonance bump in the physical regime.   However, at finite 
$\Delta_{\rm max}$, the Pad\'e approximation procedure works much more efficiently if we fit the Breit-Wigner resonance and treat it separately.}   Subtracting out bound states is trivial since they appear directly in the eigenspectrum of the Hamiltonian and consequently are easily isolated individual terms in the sum over states in (\ref{eq:spectralDensityDefinition}).  Removing resonances is a more involved process.  To identify the contribution from the resonance, we first take the raw spectral density (\ref{eq:spectralDensityDefinition}) from truncation and look at the integrated spectral density ${\cal I}(s) \equiv \int_0^s ds' \rho_{T_{--}}(s')$; because of discreteness, this is a sum of step functions.  The resonance is a prominent feature where ${\cal I}(s)$ rises  suddenly and steeply.  We fit this behavior to an integrated Breit-Wigner form:
\be
{\cal I}_{\rm BW}(s) \equiv c_0 + a \left( \frac{1}{2} - \frac{\tan^{-1} \left( \frac{ M^2-s}{M \Gamma} \right)}{\pi} \right),
\label{eq:IntegratedBW}
\ee
where we obtain the parameters $a, c_0, M$ and $\Gamma$ from the fit.  Finally, we want to perform a Pad\'e approximation to the time-ordered correlator after removing the contribution from the resonance:
\be
G_{\rm BW}(s) \equiv \int_{4m_{\rm gap}^2}^\infty d \mu^2 \frac{\rho_{\rm BW}(\mu^2)}{s-\mu^2 +i \epsilon}, \qquad \rho_{\rm BW}(\mu^2) \equiv a \frac{ M \Gamma}{(s-M^2)^2 + M^2 \Gamma^2}.
\ee
To summarize, the final object that we Pad\'e approximate is the time-ordered correlator with bound states and the resonance removed:
\be
G_{\rm smooth}(s) \equiv \left( \sum_{i \ne {\rm bound}} \frac{\rho_{T_{--}}(\mu_i)}{s - \mu^2 + i\epsilon}\right) - G_{\rm BW}(s).
\ee
In practice, we have found that a $(4,4)$ Pad\'e approximant in the variable $z$ works well.  At that point, we have the spectral density decomposed into three different pieces: the bound states, which we know directly from the Hamiltonian eigenstates; the resonance, which we have in the form $\rho_{\rm BW}(\mu^2)$ with the parameters $a,M,$ and $\Gamma$ from the fit of (\ref{eq:IntegratedBW}); and the remainder, which we know from  $\rho_{T_{--}; \mathrm{ smooth}} \equiv -\frac{1}{\pi} \Im \tilde G_{\rm smooth}(s)$.  Adding these three contributions back together gives us the full spectral density.   We show the result for the spectral density $\rho_{T_{--}}$ at $m_q = 0.125$ in Fig. \ref{fig:pltSDSG}. 

\section{Conclusions and Future Directions}

Gauge theories in two dimensions are remarkably rich and capture many  important features of  their higher dimensional cousins.  The case we have  focused on in this work was motivated in particular by its similarity to real-world QCD -- with $N_c=3$, the theory is nonintegrable, and with small but nonvanishing quark mass $m_q$, there is a ``pion'' that is parametrically light compared to the confinement scale.  Moreover, the pion is well-described by a chiral Lagrangian, which is also nonintegrable due to the presence of an infinite series of effective interactions at all orders in $m_q$, but  approaches the  Sine-Gordon model at very small $m_q$.  Although still a caricature of our world, it is nevertheless a recognizable one \cite{Coleman:1985rnk}.

The major technical challenge solved in this work is that a small nonzero quark mass causes the eigenfunctions of the Hamiltonian to behave nonanalytically, $\sim x^\alpha$ with $0< \alpha <1$, at small parton-$x$. To obtain accurate numeric results, we needed a basis that accommodated this boundary behavior, in order to achieve fast convergence as the number of basis states increases, and moreover a method for efficiently computing Hamiltonian matrix elements in this basis.  The method developed here used the fact that in the free fermion UV CFT, one can treat the nonlocal operator $\partial^\alpha \psi$ as a generalized free field, and use it to construct composite primary operators and their corresponding basis states in Lightcone Conformal Truncation (LCT).  The resulting basis was remarkably efficient, not only in terms of the mass eigenstates but also in terms of how accurately they were able to capture the correct correlators in the IR.  In particular, as shown in Fig.~\ref{fig:pltCFuncSG}, our Pad\'e approximation of the correlators was extremely accurate, significantly better than the raw truncation spectral density would suggest is possible. 

Despite the focus here on one particular model, namely that of an $SU(3)$ gauge group with a single Dirac fermion in the fundamental representation, perhaps the most interesting application of these methods is to the vast range of IR CFTs that can be obtained with vanishing quark masses by varying the gauge group and matter content. The IR has been conjectured  \cite{Delmastro:2021otj,Kutasov:1993gq, Kutasov:1994xq,Bhanot:1993xp} to be described by gauged WZW coset models, where the specific coset is a quotient of the global symmetries of the UV free fermion theory divided by the gauge group.  Gauged WZW coset models are extremely versatile and can be used to construct many well-known CFTs, including minimal models and much more. By turning on small quark masses, one can also study relevant deformations of such models, as we did in this work for the compact free boson.  %{\red say more?}  

In fact, if one is just interested in the dynamics of the light sector when $m_q$ is small, it seems very likely that one could simplify and improve the method used here by taking $m_q \rightarrow 0$ and $g \rightarrow \infty$ while keeping the physical light mass $m_\pi \sim \sqrt{m_q g}$ fixed. The results in this work with very small numeric values of $m_q$ indicate that this is a well-behaved limit, and so it would be interesting and useful to work out the rules for basis states and their Hamiltonian matrix elements in this limit directly.

\begin{center}
\subsection*{Acknowledgments}
\end{center}

We thank  Zuhair Khandker, Silviu Pufu, Matthew Walters, and Xi Yin for helpful conversations, and Matthew Walters for collaboration at an early stage of the project; we also thank Igor Klebanov and Matthew Walters for comments on an earlier draft. ALF and EK were supported in part by the US Department of Energy Office of Science under Award Number DE-SC0015845, NA, ALF and EK were supported in part by  the Simons Collaboration Grant on the Non-Perturbative Bootstrap, and ALF in part by a Sloan Foundation fellowship. YX was supported by a  Yale Mossman  Prize Fellowship in Physics.

\

\appendix

\section{$H_{\rm eff}$ Derivation of Induced Gauge Interaction}
\label{app:GaugeHeff}

The correct prescription for integrating out the gauge field in lightcone gauge is somewhat subtle.  In 't Hooft's original paper \cite{t1993two} on his solution to the infinite $N_c$ limit of the theory, he derived the `principal value' prescription (\ref{eq:PV}) by solving the Bethe-Salpeter equations that emerged from an analysis of Feynman diagrams, using a hard IR cutoff on momentum $|k| > |k_{\rm min}|$.  He found that all physical quantities remained finite  in the limit $k_{\rm min} \rightarrow 0$, and his principal value prescription emerged naturally.  Nevertheless, one might wonder whether this prescription is still the correct one at finite $N_c$, and in fact some of the literature that followed debated the issue.  In any case, in this appendix we will take the opportunity to show how 't Hooft's principal value prescription emerges from the procedure in \cite{Fitzpatrick:2018ttk} for an effective lightcone Hamiltonian $H_{\rm eff}$ that results from integrating out zero modes that are discarded in lightcone quantization.  

The basic idea of the effective Hamiltonian $H_{\rm eff}$ is that it follows from matching the time-dependence of correlators computed in equal-time and lightcone quantization, since the correlators are physical and should agree in either quantization scheme.  In particular, given the correlators, one can extract the Hamiltonian matrix elements in either quantization scheme by using the relation $\< \CO, t_i | H |  \CO', t_i\> = \lim_{t_f\rightarrow t_i} i \partial_{t_f} \< \CO, t_i | \CO', t_f\>$.   By using the Dyson series for the unitary time evolution operator $U(t,0) = 1 - i \int_0^t dt_1 H(t_1) -\frac{1}{2} \int_0^t dt_1 dt_2 {\cal T}\{ H(t_1) H(t_2) \} + \dots$, one can explicitly see that $\delta(x^+)$ terms in correlators can produce terms that show up in the lightcone Hamiltonian but not the equal-time Hamiltonian (because eg they collapse the double integral in the second order term in the Dyson series to a single integral, but only in lightcone quantization).  See \cite{Fitzpatrick:2018ttk}, or appendix B of \cite{Anand:2020gnn}, for more details.  

With this prescription in hand, the key point is to look at the gauge boson propagator for $A_+$ in equal-time quantization in lightcone gauge:
\be
G_{\rm ET}(x) = \frac{1}{2\pi}\frac{x^-}{-x^+ +  i \epsilon {\rm sgn}( x^-)} = \frac{1}{2\pi} \left( - P.V. \frac{x^-}{x^+}  - i \pi \delta(x^+) |x^-| \right) .
\ee
When we integrate out $A_+$, the term that must be added to the effective lightcone Hamiltonian $H_{\rm eff}$ in order to match correlators computed in equal-time quantization is the coefficient of the $\delta(x^+)$ term in the gauge boson propagator.  That is, we should interpret $\frac{1}{\partial^2}$ in position space as $f(x) \frac{1}{\partial^2} g(x) = - \frac{i}{2}  \int dy^- f(x) |x^- - y^-| g(y)$.  It is then straightforward to Fourier transform $|x^-|$ and see that this rule corresponds in momentum space to 't Hooft's principal value prescription (\ref{eq:PV}).

\section{Details of Modified Basis}
\label{app:ModifiedBasis}

    We will start from the building blocks $\d^\alpha\psi$, with dimension $\alpha+\frac{1}{2}$. When using double-trace construction
    \begin{align}
        A \overset{\leftrightarrow}{\d^\ell} B
        \equiv \sum_{k=0}^\ell C^\ell_k (\Delta_A,\Delta_B) \d^{\ell-k}A \d^{k} B \, ,
    \end{align}
    we modify the dimensions $\Delta_A$ and $\Delta_B$ that go into the coefficients
    \begin{align}
        C^\ell_k (\Delta_A,\Delta_B) \equiv \frac{(-1)^k \Gamma \left(\ell +2 \Delta _A\right) \Gamma \left(\ell +2 \Delta _B\right)}{k! (\ell
       -k)! \Gamma \left(k+2 \Delta _B\right) \Gamma \left(-k+\ell +2 \Delta _A\right)}\, ,
    \end{align}
    where
    \begin{align}
    \Delta_A, \Delta_B = \text{(degree)} + n \times \left(\alpha+\frac{1}{2}\right) \, .
    \end{align}
    For example, a neutral two fermion operator in QCD with degree $\ell$ can be expressed as
    \begin{align}
    \label{eq:alternativeBCJacobiPolynomial}
    [\d^\alpha\psi^\dagger \d^\alpha\psi]_\ell &\equiv
    \d^\alpha\psi^\dagger \widehat{P}_{\ell}^{(2\alpha,2\alpha)}
    \left(
    \overrightarrow{\d} - \overleftarrow{\d}
    \right) \d^\alpha\psi \nn \\
    &= \mu^{(2\alpha,2\alpha)}
    \sum_k 
    C^\ell_k (\alpha+1/2,\alpha+1/2)
    \d^{\alpha+\ell-k}\psi^\dagger
    \d^{\alpha+k}\psi \, .
    \end{align}
    In both radial quantization and Wick contraction (\ref{eq:alternativeBCJacobiPolynomial}) should give orthogonal states. This is the basis of the Generalized Free Field theory.

    \begin{figure}[htbp]
    \centering
    \includegraphics[width=.45\linewidth]{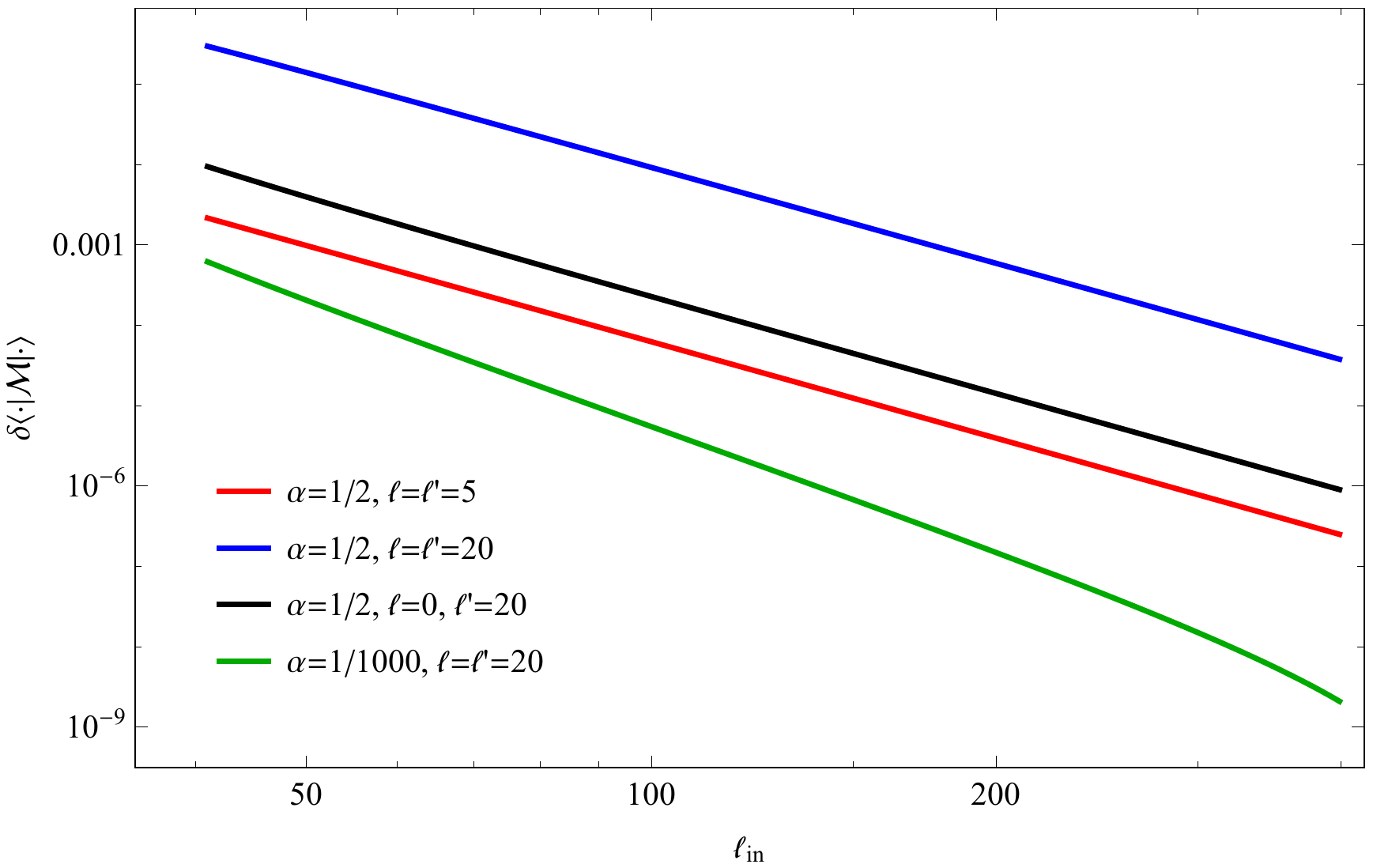}
    \caption{\label{fig:plotGFFQCDIntElemAccuracy}
    The consecutive difference between individual matrix elements at $\ell_{\rm in}$ and $(\ell_{\rm in}+2)$, as a function of $\ell_{\rm in}$. The matrix elements converge as power law. Small $\alpha$ tends to have better accuracy. The accuracy of low $\ell$ is better than higher $\ell$. 
    }
    \end{figure}
    In computing the interaction term matrix elements with respect to the large $N_c$ GFF basis, we encounter the integral
    \begin{align}
    \label{eq:QCDIntElemGFFBasis}
    \< \widehat{P}_{\ell}^{(2\alpha,2\alpha)}| \CM | \widehat{P}_{\ell'}^{(2\alpha,2\alpha)} \>
    \equiv 
    \mathcal{P}
     &\int dx dx' \, 
    \bigg[
    x^{\alpha}
    (1-x)^{\alpha}
    x'^{\alpha}
    (1-x')^{\alpha}
    \widehat{P}_{\ell}^{(2\alpha,2\alpha)}(1-2x)
    \widehat{P}_{\ell'}^{(2\alpha,2\alpha)}(1-2x')
    \nn\\
    &-x^\alpha (1-x)^\alpha
    \widehat{P}_{\ell}^{(2\alpha,2\alpha)}(1-2x)
    \widehat{P}_{\ell'}^{(2\alpha,2\alpha)}(1-2x)
    \bigg]
    \frac{1}{(x-x')^2}
      \, ,
    \end{align}
    and we do not have its closed-form expression. We can approximately compute them by separating the wave function into pieces
    \begin{align}
    \label{eq:expandingGFFBasisAsLegendre}
    x^{\alpha}(1-x)^{\alpha}
    \widehat{P}_{\ell}^{(2\alpha,2\alpha)}(1-2x)
    = &\, \widehat{P}_{\ell}^{(2\alpha,2\alpha)}(1) x^{\alpha}
    + \widehat{P}_{\ell}^{(2\alpha,2\alpha)}(-1) (1-x)^{\alpha} \nn \\
    &+ \sum_k^\infty c_\ell^k  \widehat{P}_{k}^{(0,0)}(1-2x)
    \end{align}
    where the first two pieces are monomials that carry the boundary condition. The remaining piece has trivial boundary condition, which we expand as Legendre polynomials with coefficients
    \begin{align}
    c_\ell^k &\equiv 
    \int_0^1 dx\,   x^{\alpha}(1-x)^{\alpha}
    \widehat{P}_{\ell}^{(2\alpha,2\alpha)}(1-2x) 
     \widehat{P}_{k}^{(0,0)}(1-2x)  \nn \\
    &~~~~~- \widehat{P}_{\ell}^{(2\alpha,2\alpha)}(1) \,
    x^{\alpha} 
     \widehat{P}_{k}^{(0,0)}(1-2x) \nn \\
    &~~~~~- \widehat{P}_{\ell}^{(2\alpha,2\alpha)}(-1) \,
     (1-x)^{\alpha}
     \widehat{P}_{k}^{(0,0)}(1-2x) \, ,
    \end{align}
    where the boundary-to-Lagendre term has closed-form expression
    \begin{align}
    \int_0^1 dx\,
    x^{\alpha}
     \widehat{P}_{k}^{(0,0)}(1-2x)
     &= \frac{\sqrt{2 k+1} \left(-\alpha\right)_k}{\left(\alpha+1\right)_{k+1}}
    \end{align}
    and the Jacobi-to-Legendre term is computed as a sum over pairs of monomials
    \begin{align}
    \int_0^1 dx\,
    x^{m_1+m_2 }
    x^{\alpha}(1-x)^{\alpha}
    =
    \frac{\Gamma \left(\alpha+1\right) \Gamma \left(\alpha+m_1+m_2+1\right)}{\Gamma \left(2\alpha +m_1+m_2+2\right)}
    \end{align}
    weighted by the coefficients in the Jacobi and Legendre polynomials.

    Now we split the matrix element (\ref{eq:QCDIntElemGFFBasis}) in terms of the pieces in (\ref{eq:expandingGFFBasisAsLegendre}). We gather like terms using the symmetry of Jacobi polynomials, and the final result is
    \begin{align}
    \< \widehat{P}_{\ell}^{(2\alpha,2\alpha)}| \CM | \widehat{P}_{\ell'}^{(2\alpha,2\alpha)} \>
    &= \pr{
        \widehat{P}_{\ell}^{(2\alpha,2\alpha)}(1)
        \widehat{P}_{\ell'}^{(2\alpha,2\alpha)}(1)
        +\widehat{P}_{\ell}^{(2\alpha,2\alpha)}(-1)
        \widehat{P}_{\ell'}^{(2\alpha,2\alpha)}(-1)
    } \< x^{\alpha} | \CM | x^{\alpha} \> 
    \nn \\
    &~~~ +\pr{
        \widehat{P}_{\ell}^{(2\alpha,2\alpha)}(1)
        \widehat{P}_{\ell'}^{(2\alpha,2\alpha)}(-1)
        +\widehat{P}_{\ell}^{(2\alpha,2\alpha)}(-1)
        \widehat{P}_{\ell'}^{(2\alpha,2\alpha)}(1)
    } \< x^{\alpha} | \CM | (1-x)^{\alpha} \>
    \nn \\
    &~~~ + \sum_{k'} c_{\ell'}^{k'} \pr{
        \widehat{P}_{\ell}^{(2\alpha,2\alpha)}(1)
        + (-1)^{k'} \widehat{P}_{\ell}^{(2\alpha,2\alpha)}(-1)
    } \< x^{\alpha} | \CM | \widehat{P}_{k'}^{(0,0)} \>
    \nn \\
    &~~~ + \sum_k c_{\ell}^{k} \pr{
        \widehat{P}_{\ell'}^{(2\alpha,2\alpha)}(1)
        + (-1)^{k} \widehat{P}_{\ell'}^{(2\alpha,2\alpha)}(-1)
    } \< x^{\alpha} | \CM | \widehat{P}_{k}^{(0,0)} \>
    \nn \\
    &~~~ + \sum_k \sum_{k'} c_{\ell}^{k} c_{\ell'}^{k'} 
    \< \widehat{P}_{k}^{(0,0)} | \CM | \widehat{P}_{k'}^{(0,0)} \>
      \, .
    \end{align}
    All integrals that show up in the above formula have closed-form expression
    \begin{align}
    \CM_{k,k'}\equiv& \,
    \< \widehat{P}_{k}^{(0,0)} | \CM | \widehat{P}_{k'}^{(0,0)} \>
    \nn \\
    =& \, \mathcal{P} \int dx dx' \,
    \frac{\widehat{P}_{\ell}^{(0,0)}(1-2x)
    \widehat{P}_{\ell'}^{(0,0)}(1-2x')
    - \widehat{P}_{\ell}^{(0,0)}(1-2x)
    \widehat{P}_{\ell'}^{(0,0)}(1-2x)
    }{(x-x')^2} \nn \\
    =& \,
    -H_{ \left(k+k'\right)/2}-H_{
       \left(-k_m+k_M-1\right)/2}+
       H_{\left(k_M-1\right)/2}+H_{k_M/2} \\
    I_1(a,b) \equiv& \,
    \< x^a | \CM | x^b \> = \mathcal{P}
     \int dx dx' \, \frac{
            x^a x'^b - x^{a+b}
        }{(x-x')^2} \nn \\
       =& \, \frac{a H_a+b H_b-1}{a+b} - H_{a+b-1} \, 
    \\
    I_1(0,0) =& \, 0 \\
    I_2(a,b) \equiv& \,
    \< x^a | \CM | (1-x)^b \> = 
    \mathcal{P}\int dx dx' \, \frac{
        x^a (1-x')^b 
        - x^a (1-x)^b 
    }{(x-x')^2} 
    \nn \\
    =&\,\frac{1}
       {2 a (a+1) \Gamma (a+b+1)}  \nn \\
    &~~~~\times\bigg(
    -2 a (a+1) 
     \Gamma (b+1) \Gamma (a+b+2) 
    -2 \Gamma
       (a+2) \Gamma (b) (a b \psi ^{(0)}(b)+a+b)
     \nn \\
    &~~~~~~~~ +2 \Gamma (a+b+1)
     \left.\frac{\d}{\d t}\left(
     \, _3\tilde{F}_2(2,-a,b+1;b+2,t;1)
     \right)\right|_{t=2}
    \bigg) \\
    I_2(0,b) =& \, I_2(a,0) = 0 \, 
    \end{align}
    where $\psi^{(0)}(b)$ is the digamma function, $\, _3\tilde{F}_2$ is the regularized hypergeometric function, and $H_n$ is the $n$-the harmonic number.

    In practice we need to truncate the infinite sums over $k$ and $k'$ to some ``internal cutoff'' $\ell_{in}$. 
    We study the convergence of individual matrix elements in Figure \ref{fig:plotGFFQCDIntElemAccuracy}. 
    We see that the matrix elements all converge quickly to the exact value. At $\dmax = 21+2\alpha$, we can compute the matrix elements to $\ell_{\rm in} = 400$ in a few minutes, and the top matrix element have about 4 digits of accuracy. The fact that low $\ell$ matrix elements converge faster than high $\ell$, and that higher $\ell$ basis states decouple from the low-lying spectrum indicate that the precision of the eigenvalues may be better. We conclude that at $\dmax = 21+2\alpha$ and $\ell_{\rm in} = 400$ the Hamiltonian is reliable up to a small numerical error.

    The fermion mass term has closed-form expression 
    \begin{align}
    \< \widehat{P}_{\ell}^{(2\alpha,2\alpha)} | \CM^{\rm(mass)} | \widehat{P}_{\ell'}^{(2\alpha,2\alpha)} \>
    &= \int dx \,
    x^{\alpha-1} (1-x)^{\alpha-1}
    \widehat{P}_{\ell}^{(2\alpha,2\alpha)}(1-2x)
    \widehat{P}_{\ell'}^{(2\alpha,2\alpha)}(1-2x)
    \nn \\
    &= \begin{cases}
    \frac{2 \Gamma \left(2\alpha +\ell _m+1\right) \Gamma \left(2\alpha +\ell
       _M+1\right)}{2\alpha  \Gamma \left(\ell _m+1\right) \Gamma \left(4 \alpha +\ell
       _M+1\right)} & \ell = \ell' {\rm ~mod~} 2 \\
    0 & {\rm otherwise}
    \end{cases}
     \, ,
    \end{align}
    where $\ell_M \equiv \max(\ell,\ell')$ and $\ell_m \equiv \min(\ell,\ell')$.

    The above algorithm computes the matrix elements at large-$\Nc$, and can be easily generalized to the finite-$\Nc$ case. At finite-$\Nc$, we encounter matrix elements with generic particle numbers. Following the procedure of \cite{Anand:2020gnn}, we arrive at an expression
    \begin{equation}
    \begin{aligned}
        \frac{\Mcal_{\Kvec \Kvec'}}{2p(2\pi)\delta(p-p')} = \sum_{a,b,a',b'}  \frac{
            2\pi^2 
            p^{\Delta+\Delta^\prime-1}
            \tilde A_{\kvec,\kvec^\prime}^{(a,b,a',b')}
        }{\Gamma(\Delta+\Delta^\prime-1)} I(a,b,a^\prime,b^\prime) \, ,
        \label{eq:gaugesummaryeqFirst}
    \end{aligned}
    \end{equation}
    where $A_{\kvec,\kvec^\prime}^{(a,b,a',b')}$ comes from spectators and $I(a,b,a^\prime,b^\prime)$ comes from the active part. With the $\d^\alpha\psi$ basis, the powers $a,b,a',b'$ are no longer integers but integers shifted uniformly by $\alpha$. For the spectator factor $A_{\kvec,\kvec^\prime}^{(a,b,a',b')}$ and many cases in $I(a,b,a^\prime,b^\prime)$, the expression only contains Gamma functions, and the continuation to non-integers is straightforward. For most the generic $I(a,b,a^\prime,b^\prime)$, we need to compute 
    \begin{equation}
    \begin{aligned}\label{eq:gauge-interaction-momentum}
    I(a,b,a^\prime,b^\prime) =& \frac{
    \Gamma(\Delta+\Delta^\prime-1)
    }
    {
    \Gamma(a)\Gamma(b)\Gamma(a^\prime)\Gamma(b^\prime)
    \Gamma(c)}
    \int \, dx_1 x_1^{a+b+a^\prime+b^\prime-4}(1-x_1)^{c-1}  \\
    &~~\times 
    \int dx_2 dx_3 \, 
    \frac{x_2^{a-1} (1-x_2)^{a^\prime-1} \left( x_3^{b-1} (1-x_3)^{b^\prime-1} - x_2^{b-1} (1-x_2)^{b^\prime-1}\right) }{(x_2-x_3)^2} \, .
    \end{aligned}
    \end{equation}
Directly evaluating   (\ref{eq:gauge-interaction-momentum}) by brute force for non-integer powers of $x_2, x_3$ can be computationally expensive.
 However, notice that the Jacobi polynomials $x^{\alpha}(1-x)^{\alpha}
    \widehat{P}_{\ell}^{(2s,2s)}(1-2x)$ form a complete basis of polynomials on $x\in[0,1]$ with boundary condition $x^{\alpha}(1-x)^{\alpha}$. 
    So all possible $x_2,x_3$ integrals in (\ref{eq:gauge-interaction-momentum}) can be written as a linear combination of (\ref{eq:QCDIntElemGFFBasis}), and there is actually nothing new to compute.

\section{Form factors in Sine-Gordon theory}
    \label{sec:FormFactors}
    In this appendix we show the details in the bound state form factors in (\ref{eq:BoundStateFormFactors}).
    We comput the $c$-function from the spectral density of the stress tensor  
    \begin{equation}
    c(\mu) = 12\pi \int_0^{\mu^2} ds \frac{\rho_{\Theta}(s)}{s^2} \, .
    \end{equation}
    We focus on the contribution from the $k$-th bound state $b_k$. The spectral density of one-particle states is $\rho_k = |F_{b_k}^{\Theta}|^2 \delta(s-m_k)$, and $m_k = 2 m_s \left(\sin\frac{\pi}{2}k\nu\right)^4$, so we have
    \begin{equation}
    c_k = \frac{12\pi |F_{b_k}^{\Theta}|^2}{\left(2 \sin\frac{\pi}{2}k\nu\right)^4}
    \end{equation}
    where we set the soliton mass $m_s = 1$. The form factors of bound states $b_k$ can be extracted from the pole of the $\left\< s \bar s | \Theta | 0 \right\>$ form factor, divided by the residue of the pole of the bound state in the $b_1 b_2 \rightarrow b_1 b_2$ S-matrix \cite{Babujian:1998uw}:
    \begin{equation}
    F_{b_k}^{\Theta} = \underset{\theta\rightarrow\theta_{k}}{\rm Res} F_{s\bar s}^{\Theta} \times 
        \sqrt{2} \left( 2 i \underset{\theta\rightarrow\theta_{k}}{\rm Res} S_{+}\right)^{-\frac{1}{2}} \, ,
    \end{equation}
    where the bound state $b_k$ is associated with a pole at $\theta_k = i \pi (1-k\nu)$. 
    The parity-even $s \bar s$-S-matrix has an integral expression
    \begin{equation}
    S_{+} = \frac{\sinh \frac{i \pi }{\nu }+\sinh \frac{i \pi  x}{\nu }}{\sinh \frac{i \pi  (1-x)}{\nu }} \int \frac{dt}{t}\frac{\sinh \left(\frac{1}{2} (1-\nu ) t\right) \sinh (t x)}{ \cosh \frac{t}{2} \sinh \frac{\nu  t}{2} } \, .
    \end{equation}
    The formulas for $F_{s\bar s}^{\Theta}$ can be found in \cite{Babujian:2001xn}:
    \begin{equation}
    F_{s\bar s}^{\Theta} (\theta) = 
    \frac{2\sqrt{2}i}{\nu} 
    \sinh\left(\frac{\theta}{2}\right)
    \frac{\sinh\frac{1}{2}(i\pi-\theta)}{\sinh\frac{1}{2\nu}(i\pi-\theta)} f_{s\bar s}^{\rm min}(\theta)\, ,
    \end{equation}
    where 
    \begin{equation}
    f^{\rm min}_{s \bar s} (\theta) = 
    \frac{dt}{t}
    \frac{\sinh \left( \frac{1}{2} (1-\nu ) t \right) (1-\cosh t (1-x))}{2 \sinh t \cosh \frac{t}{2} \sinh \frac{\nu  t}{2}} \, .
    \end{equation}
    One can check that $f^{\rm min}_{s \bar s}$ is regular at $\theta\rightarrow\theta_k$, so the residue is 
    \begin{equation}
    {\rm Res}F_{s\bar s}^{\Theta} (\theta) = -2\sqrt{2} i^{k+1} \sin (\pi  k \nu ) f^{\rm min}_{s \bar s} (\theta) \, .
    \end{equation}
    The stress tensor has support at even $k$. At $\Nc = 3$, $\nu = \frac{1}{5}$, and we have $c_2$ and $c_4$ contribution to the $c$-function.

\bibliographystyle{utphys}
\bibliography{RefYuan}

\end{document}